\documentclass[twocolumn]{aastex631}

\shorttitle{Towards UMP DLAs}
\shortauthors{Welsh et al.}
\graphicspath{{./}{figures/}}
\newcommand{\jqso}{J0903$+$2628}
\newcommand{\fe}[1]{Fe\,{\sc ii}~$\lambda#1$}
\newcommand{\oi}[1]{O\,{\sc i}~$\lambda#1$}
\newcommand{\al}[1]{Al\,{\sc ii}~$\lambda#1$}

\newcommand{\kms}{km\,s$^{-1}$}

\defcitealias{HegerWoosley2010}{HW10}
\defcitealias{Cooke2017}{C17}

\begin{document}

\title{Towards ultra metal-poor DLAs: linking the chemistry of the most metal-poor DLA to the first stars}

\correspondingauthor{Louise Welsh}
\email{louise.welsh@unimib.it}

\author[0000-0003-3174-7054]{Louise Welsh}
\affiliation{Dipartimento di Fisica G. Occhialini, Universit\`a degli Studi di Milano Bicocca, Piazza della Scienza 3, I-20126 Milano, Italy}
\affiliation{INAF – Osservatorio Astronomico di Brera, via Bianchi 46, I-23087 Merate (LC), Italy}

\author[0000-0001-7653-5827]{Ryan Cooke}
\affiliation{Centre for Extragalactic Astronomy, Department of Physics, Durham University, South Road, Durham DH1 3LE, UK}

\author[0000-0001-6676-3842]{Michele Fumagalli}
\affiliation{Dipartimento di Fisica G. Occhialini, Universit\`a degli Studi di Milano Bicocca, Piazza della Scienza 3, I-20126 Milano, Italy}

\affiliation{INAF - Osservatorio Astronomico di Trieste, via G. B. Tiepolo 11, I-34143 Trieste, Italy}

\author[0000-0002-5139-4359]{Max Pettini}
\affiliation{Institute of Astronomy, University of Cambridge, Madingley Road, Cambridge CB3 0HA, UK}



\begin{abstract}
We present new Keck/HIRES data of the most metal-poor damped Ly$\alpha$ (DLA) system currently known. By targeting the strongest accessible Fe\,{\sc ii} features, we have improved the upper limit of the [Fe/H] abundance determination by $\sim1$~dex, finding [Fe/H]~$<-3.66~(2\sigma)$. We also provide the first upper limit on the relative abundance of an odd-atomic number element for this system [Al/H]~$<-3.82~(2\sigma)$.  Our analysis thus confirms that this $z_{\rm abs}\simeq 3.07$ DLA is not only the most metal-poor DLA but also the most iron-poor DLA currently known.
We use the chemistry of this DLA, combined with a stochastic chemical enrichment model, to probe its enrichment history. We find that this DLA is best modelled by the yields of an individual Population III progenitor rather than multiple Population III stars. We then draw comparisons with other relic environments and, particularly, the stars within nearby ultra-faint dwarf galaxies. We identify a star within Bo\"{o}tes~I, with a similar chemistry to that of the DLA presented here, suggesting that it may have been born in a gas cloud that had similar properties. The extremely metal-poor DLA at redshift $z_{\rm abs}\simeq3.07$ (i.e. $\sim 2$ Gyrs after the Big Bang) may reside in one of the least polluted environments in the early Universe.

\end{abstract}



\section{Introduction} \label{sec:intro}
Tracing chemical evolution from the Cosmic Dawn at $z\sim15-30$ to the present epoch will reveal how the Universe transformed from primordial gas (i.e. primarily hydrogen and helium) to the complex composition of chemical elements we observe locally. Our current understanding suggests that the first generation of stars (known as Population III, or Pop III stars) were the catalysts of this process in the early Universe \citep[see e.g.][]{BrommYoshida2011}. Since the properties of these stars remain elusive to observations, the relative quantities of metals that they produced is also an open question. To understand the earliest epochs of chemical enrichment, we must uncover the properties of the first stars. 

Observationally, we have searched for these stars in the local Universe for over four decades \citep[e.g.][]{Bond1980, Beers1985, Ryan1991, Beers1992, McWilliam1995, Ryan1996, Cayrel2004, Beers2008, Christlieb2008, Roederer2014, Howes2016, Starkenburg2017}. None have been found. 

Cosmological hydrodynamic simulations that follow the formation of Pop III stars from cosmological initial conditions suggest that the first stars were more massive than the Sun, with typical masses between $10 < M/{\rm M_{\odot}} <100$ \citep{Tegmark1997, BarkanaLoeb2001, Abel2002, BrommCoppiLarson2002, Turk2009, Greif2010, Clark2011, Hirano2014, Stacy2016}. If Pop III stars really were limited to these mass scales, their short lifetimes mean they may only be visible at very high redshift; directly observing the supernovae of these stars is one of the flagship goals of the recently launched James Webb Space Telescope (JWST). This facility provides a new avenue to observationally investigate the first stars and galaxies which may revolutionise this field in the coming years \citep{Gardner2009}. A complementary approach to searching for these stars directly is to search for the elements that they produced.

Reservoirs of gas, detected as absorption along the line-of-sight towards unrelated background quasars, are reliable probes of cosmic chemical evolution. These low density structures allow us to trace the evolution of metals from $0<z<5$ with a consistent degree of sensitivity \citep{Peroux2020}. The absorption line systems whose column density of neutral hydrogen exceeds $\log_{10} (N({\rm H}$\,{\sc i}$)/{\rm cm}^{-2})\ge 20.3$ are known as Damped Ly$\alpha$ systems (DLAs; see \citealt{Wolfe2005} for a review). DLAs are uniquely suited to high-precision chemical abundance studies; the large column density of neutral hydrogen means that the gas is optically thick to ionizing radiation and the constituent metals reside primarily in a single, dominant ionization state. This negates the need for ionization corrections and, thus, the column density of the observed ionic species can be used to accurately determine the relative metal abundances of the gas reservoir. These objects are most easily studied in the redshift interval $2<z<3$, when the rest-frame UV lines (e.g., H\,{\sc i}~$\lambda1215$ and \fe{2382}) are shifted into the optical wavelength range. However, DLAs can be studied at any redshift once an appropriate sightline (e.g., a quasar) has been identified. 

Studies which probe $z>6$ sightlines trace the composition of gas closer to the epoch of the Cosmic Dawn \citep{Becker2011, Becker2019, Bosman2022, Dodorico2022, Dodorico2023}. However, at these redshifts, it can be challenging to observe the characteristic Ly$\alpha$ absorption, which is required to determine the metallicity. The column density of neutral hydrogen can only be determined for proximate systems which are less frequent and suffer additional challenges due to the proximity to the background quasar \citep{Simcoe2012, Banados2019}. It may be that the systems found at lower redshift offer the best opportunity to identify relic metals with high-precision.

Indeed, there may be some near-pristine gaseous reservoirs in the early Universe that have been solely enriched by the supernovae (SNe) of the first stars \citep{Erni2006, Pettini2008, Cooke2017, Welsh2019, HazeNunez2022}. The metallicity of these reservoirs provides an indication of the level of enrichment the gas has experienced from stellar populations. In principle, this property is easily defined as the amount of metals in the system relative to the hydrogen content. This is often expressed as [X/H]\footnote{This denotes the logarithmic number abundance ratio of elements X and H relative to their solar values X$_{\odot}$ and H$_{\odot}$, i.e. $[{\rm X / H}] =  \log_{10}( N_{{\rm X}}/N_{{\rm H}}) - \log_{10} (N_{{\rm X}}/N_{{\rm H}})_{\odot}$.}. However, the metal (or combination of metals) used in this expression is dependent on the system being analysed and, ultimately, on the accessible features. 

A typical metallicity tracer is often [Fe/H] (i.e. the relative abundance of iron to hydrogen --- which we call the Fe-metallicity); anything that is 1000 times more Fe-poor than the Sun (i.e., [Fe/H]~$<-3$) is considered extremely metal-poor (EMP), while an environment that is 10\,000 times more Fe-poor than the Sun (i.e., [Fe/H]~$<-4$) is considered ultra metal-poor (UMP) \citep[see][for the definitions used to describe different levels of metal-paucity]{Beers2005}. As we continue to discover increasingly Fe-poor objects, peculiar chemical abundance patterns become more apparent. The lowest Fe-metallicity environments are often associated with enhanced abundances of $\alpha$-elements. This is well-documented through the analysis of metal-poor Milky Way halo stars (i.e., the stellar relics) where a deficit of Fe is frequently identified alongside an overabundance of C. These stars are known as carbon-enhanced metal-poor stars \citep[i.e., CEMP stars; see][]{Beers2005}. There have been reports of a similar carbon-enhancement in both metal-poor DLAs \citep{Cooke2011a} and lower column density gas reservoirs \citep[i.e., Lyman limit systems; LLSs --- see][]{zou2020, Saccardi2023}. 
Though, an updated analysis of the C-enhanced DLA revealed a [C/Fe] abundance ratio that is only modestly enhanced relative to the metal-poor DLA population \citep{Dutta2014, Welsh2020}. Generally, we note that, when $\alpha$-elements are detected in excess alongside a minimal contribution from Fe, metallicity estimates based on Fe fail to describe the overall metal-paucity of the system in question. This is worth considering when we are searching for chemically near-pristine gas. 

We are yet to detect a chemically pristine DLA, although there are multiple LLSs that appear to be entirely untouched by the process of star formation \citep[e.g.][]{Fumagalli2012,Robert2019}. There is the potential detection of an UMP DLA at $z\sim7$ \citep{Simcoe2012}, but this hinges on the modelling of the quasar emission \citep[see][for an alternative interpretation of these data]{Bosman2015}. Hence, we are also yet to firmly detect an UMP DLA. Prior to the collection of the data presented here, there have been three  high-precision detections of EMP DLAs with associated iron abundance errors $<0.20$ dex \citep{Ellison2010, Cooke2011b, Cooke2016, Welsh2022}. \citet{Rafelski2012} report a metallicity floor across a statistical sample of DLAs that is [M/H]~$\simeq-3$. Thus, the detection of an UMP DLA would mark an exceptional environment. Future surveys like WEAVE, DESI, and 4MOST will provide much improved statistics, necessary to reveal where UMP DLAs lie within the population of known absorbers \citep{Dalton2012, deJong2012, DESI2016, Pieri2016}. 

The DLA found towards the quasar SDSS J090333.55$+$262836.3 (hereafter \jqso) is the \emph{most} metal-poor DLA currently known \citep{Cooke2017}. This was determined through the analysis of absorption from low ionic species such as C\,{\sc ii}, O\,{\sc i}, and Si\,{\sc ii}. The strongest Fe\,{\sc ii} feature available with those data was the weak \fe{1260} transition; in order to observe the absorption due to the lighter atomic number elements together with the associated H\,{\sc i}, \citet{Cooke2017} adopted an instrument setup that did not cover several stronger Fe\,{\sc ii} transitions. As a result, the data presented in \citet{Cooke2017} only provided a relatively weak upper limit on the iron abundance, [Fe/H]~$<-2.81~(2\sigma)$. Given the extreme paucity of both $\alpha$-elements and Fe observed for this system, we have sought additional high resolution observations that target the stronger \fe{1608} and \fe{2382} features. These observations also target the \al{1670} feature --- providing the first insight into the abundance of an odd atomic number element for this DLA. In this paper, we utilise these additional data to reevaluate the most likely enrichment scenario of this DLA. We also present the first stochastic chemical enrichment analysis of this near-pristine gas cloud, and test if this environment might have been enriched by the first generation of stars. 

 This paper is organised as follows. Section \ref{sec:obs} describes our observations and data reduction. In Section \ref{sec:ana}, we present our data and investigate the chemical enrichment history of this system. In Section \ref{sec:disc}, we present our discussion and draw comparisons with other relic environments, before drawing overall conclusions and suggesting future work in Section \ref{sec:conc}.

\begin{figure*}
    \centering
    \includegraphics[width=0.9\textwidth]{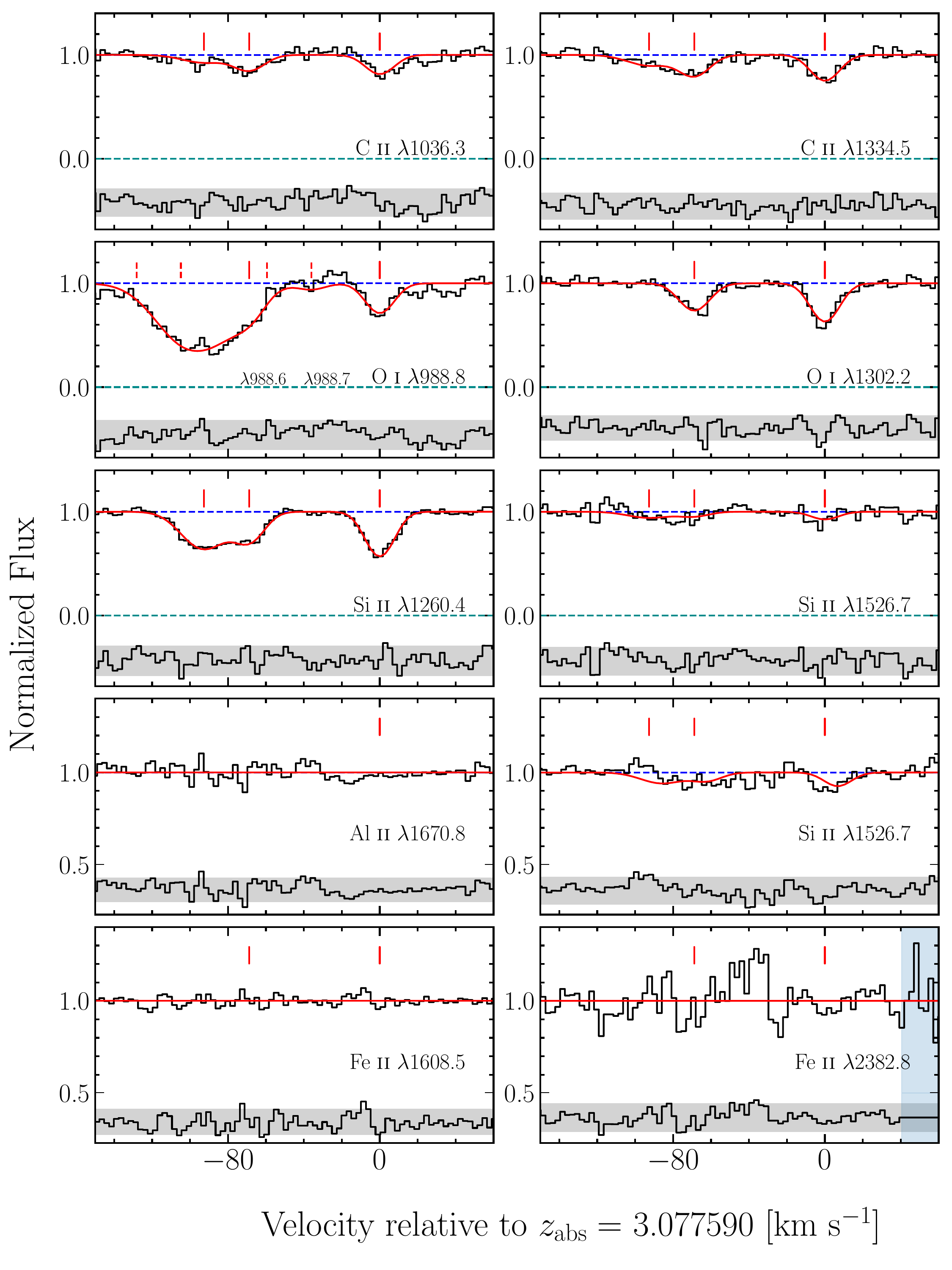}
    \caption{Continuum normalised HIRES data (black histograms) of the absorption features produced by metal ions associated with the DLA at $z_{\rm abs}=3.077590$ towards the quasar \jqso. The best-fitting model is shown with the red curves. The blue dashed line indicates the position of the continuum while the green dashed line indicates the zero-level. Note the different scale of the y-axes of the bottom two rows. These bottom two rows correspond to the data collected in 2021 while the upper three rows correspond to the 2016 Keck/HIRES data. The red ticks above the absorption features indicate the centre of the Voigt line profiles for each identified component. The dashed-red lines in the third panel highlight the components of the \oi{988} triplet. Note that the deep broadening seen at $<-50$~\kms\ in the O\,{\sc i}~988 panel is due to unrelated blending. Below the zero-level, we show the residuals of this fit (black histogram) where the grey shaded band encompasses the $2\sigma$ deviates between the model and the data. Blue-shaded regions encompass data excluded from the fit. }
    \label{fig:j0903}
\end{figure*}

\section{Observations and data reduction}\label{sec:obs}

The DLA towards the $m_{\rm r}=19.0$ quasar \jqso\ (SDSS J090333.55$+$262836.3), at $z_{\rm em}=3.22$ has previously been identified as the most metal-poor DLA currently known. We have acquired an additional 9 hours of echelle spectroscopic data on the associated quasar using the High Resolution Echelle Spectrometer \citep[HIRES;][]{Vogt1994} mounted on the 10~m Keck I telescope. These data were collected through $9 \times 3600$~s exposures on December 18$^{\rm th}$ and December 27$^{\rm th}$ 2021. We utilise the red cross-disperser to target the strongest available Fe\,{\sc ii} features that fall at red wavelengths. The data cover $5450 - 10\,000$~\AA\ with small wavelength gaps due to both the detector mosaic and the configuration of the instrument (that results in incomplete coverage across a small number of the reddest echelle orders). We use the C1 decker resulting in a slitwidth~$=0.861''$ and a nominal resolution of $R=49,000$ corresponding to a velocity full width at half maximum $v_{\rm FWHM}=6.28~{\rm km~s}^{-1}$.

The collected data were binned $2\times 2$ during readout. 
These HIRES data were reduced with the Hires REDUX\footnote{Hires REDUX is available from: \\ \url{https://www.ucolick.org/~xavier/HIRedux/}} reduction pipeline \citep{Bernstein2015}. 
This pipeline includes the standard reduction steps of subtracting the detector bias, locating and tracing the echelle orders, flat-fielding, sky subtraction, optimally extracting the 1D spectrum, and performing a wavelength calibration. The data were converted to a vacuum and heliocentric reference frame.

 Finally, we combined the individual exposures of this DLA using {\sc uves\textunderscore popler}\footnote{{\sc uves\textunderscore popler} is available from: \\\url{https://github.com/MTMurphy77/UVES\textunderscore popler}}. This corrects for the blaze profile, and allowed us to manually mask cosmic rays and minor defects from the combined spectrum. 
When combining these data we adopt a pixel sampling of 2.5~km\,s$^{-1}$. 

These new data are analysed in the following section in combination with the original HIRES data collected in 2016 \citep[presented in ][]{Cooke2017}. These data were collected using an identical instrument setup barring the wavelength coverage, which instead spanned $3700 - 6530$~\AA. The original observations were executed as $9\times 3600$~s exposures (resulting in a combined time on source across both programmes of $18$~hours). For further details, we refer the reader to the original paper. Given both spectra, we cover the optimal features necessary to conduct a detailed investigation of the chemistry of the DLA towards \jqso.

\section{Analysis}\label{sec:ana}

 Using the Absorption LIne Software ({\sc alis}) package\footnote{{\sc alis} is available from:\\ \url{https://github.com/rcooke-ast/ALIS}.}--- which uses a $\chi$-squared minimisation procedure to find the model parameters that best describe the input data --- we simultaneously analyse the full complement of high S/N and high spectral resolution data currently available for the DLA towards \jqso. We model the absorption lines with a Voigt profile, which consists of three free parameters: a column density, a redshift, and a line broadening. We assume that all lines of comparable ionization level have the same redshift, and any absorption lines that are produced by the same ion all have the same column density and total broadening. The total broadening of the lines includes a contribution from both turbulent and thermal broadening. The turbulent broadening is assumed to be the same for all absorption features, while the thermal broadening depends inversely on the square root of the ion mass; thus, heavy elements (e.g. Fe) will exhibit absorption profiles that are intrinsically narrower than the profiles of lighter elements, (e.g. C). The intrinsic model is then convolved with the line spread function of the instrument; this results in in an additional apparent broadening of the lines. 
We note that we simultaneously fit the absorption and quasar continuum. We model the continuum around every absorption line as a low-order Legendre polynomial (typically of order 3). We assume that the zero-levels of the sky-subtracted HIRES data do not depart from zero\footnote{We visually inspected the troughs of saturated absorption features to confirm this is the case.}. 
Finally, we introduce one additional free  parameter that accounts for a relative velocity shift in the wavelength solution between the two epochs. The best-fit value ($\Delta v=4\pm2$~\kms) is driven by Si\,{\sc ii}~$\lambda$1526, which is the only transition in common between the two datasets.

The DLA towards \jqso\, is best modelled by three absorption components for C\,{\sc ii} and Si\,{\sc ii} at redshifts $z_{\rm abs}= 3.077590 \pm 0.000002$, $z_{\rm abs}= 3.076653 \pm  0.000005$, and $z_{\rm abs}= 3.076331 \pm  0.000006$. The O\,{\sc i} and Fe\,{\sc ii} features are best modelled by the two highest redshift components only. We attempted to model the intrinsically weak Al\,{\sc ii} feature solely using the highest redshift component; we instead report an upper limit. Due to the blending between the components, we cannot directly determine the temperature of the gas reservoir. We therefore assume T~$=1\times10^{4}$~K \citep[this is typical for metal-poor DLAs; see][]{CookePettiniJorgenson2015, Welsh2020, Noterdaeme2021}. Given this assumed temperature, the associated turbulent broadening of these components are best described by the Doppler parameters $b_{1}=8.8 \pm 0.3$~\kms, $b_{2}=9.6 \pm 0.5$~\kms, and $b_{3}=14.9 \pm 0.7$~\kms\ respectively. The data, along with the best fitting model, are shown in Figure~\ref{fig:j0903} while the resulting column densities and relative abundances are presented in Table~\ref{tab:j0903}. We note that, when modelling the data, we allow the relative abundances of metals to vary from component to component. We have repeated our analysis under the assumption that the relative abundances of metals (i.e., the [O/Fe] ratio) must be the same in all components; the resulting total abundances are consistent with those reported in Table~\ref{tab:j0903}. Without this assumption, the third components at $z_{\rm abs} = 3.076331$ (i.e. the bluest component) indicates an unusually high Si\,{\sc ii}/C\,{\sc ii} ratio. This could be due to several possibilities: (1) an unfortunate blend near Si\,{\sc ii}~$\lambda 1260$ (at $v \simeq -90$\kms\ in Figure~\ref{fig:j0903}), that masquerades as Si\,{\sc ii} absorption; (2) ionisation effects; or (3) a peculiar enrichment source. The reality of this Si\,{\sc ii} absorption can be tested with deeper data of the other strong Si\,{\sc ii} absorption lines (e.g. this feature may not be detected in Si\,{\sc ii}~$\lambda 1526$, but the data are of too low S/N to be confident at this stage). The composition of this third component has a negligible impact on our subsequent analysis; we consider only the chemistry of the predominantly neutral components (i.e., those traced by O\,{\sc i}). We have also repeated our analysis assuming a temperature range of $(0.5 - 2.0)\times 10^{4}$~K and found that the abundances in Table~\ref{tab:j0903} are stable against these changes. This is not surprising, since all of the detected absorption lines are weak enough to be in the linear regime of the curve of growth, and are therefore relatively insensitive to the relative amounts of Doppler, thermal, and instrumental broadening. In this metallicity regime, we do not expect our observed abundances to be impacted by any appreciable dust depletion \citep{Pettini1997, Akerman2005, Vladilo2011, Rafelski2014}.

 The reported relative abundances are consistent with those found using the original HIRES data. Notably, we have improved the upper limit on the [Fe/H] abundance determination by $\sim 1$~dex from [Fe/H]~$<-2.81~(2\sigma)$ to [Fe/H]~$\leq-3.66~(2\sigma)$. The DLA in question is therefore pushing towards the UMP regime at [Fe/H]~$<-4$.  
We also place the first upper limit on the abundance of an odd-atomic number element, [Al/H]~$\leq-3.82~(2\sigma)$, of this EMP gas cloud. 

We note that the data near the \fe{2382} feature is relatively noisy compared to the data near features at bluer wavelengths. This is, in part, due to the challenges associated with observing faint targets near $9700$~\AA\ (where this feature falls) from the ground. We have checked for possible blending with known telluric absorption and found nothing significant. This feature is the strongest Fe\,{\sc ii} line (i.e. has the largest oscillator strength) covered by our spectrum. Though we do not cover the \fe{2344} feature, we do additionally observe \fe{1260} and \fe{1608}. These lines are relatively weaker, however, they fall in regions of the spectrum that are not influenced by telluric lines. The combined information provided by these Fe\,{\sc ii} features are essential to investigate the Fe content of this DLA. 

These new data reaffirm that the DLA towards \jqso\, is the most metal-poor DLA currently known. It has the lowest C, O, Si, and Fe abundance determination of any known DLA. We also searched for absorption from higher ion stages, such as Si\,{\sc iii}, so that we can determine the ionization ratio and subsequently model the density of the gas reservoir. However, the strong Si\,{\sc iii} $\lambda$1206 feature is heavily blended with unrelated absorption. In the following sections, we investigate the chemical enrichment history of this DLA using our stochastic chemical enrichment model and the yields from simulations of Population III SNe. \\
\begin{table*}
	\centering
	\caption{Ion column densities of the DLA at $z_{\rm abs} = 3.07759$ towards the quasar \jqso. The quoted column density errors are the $1\sigma$ confidence limits while the column densities are given by log$_{10}$ $N$(X)/cm$^{-2}$. }
	\label{tab:j0903}
	\tabcolsep=0.10cm
\begin{tabular}{lcccccccc}
\hline
Ion        & Transitions & Solar$^{*}$ & Comp. 1 & Comp. 2 & Comp. 3 & Total  & [X/H] & [X/Fe] \\ 
  & used [\AA] & & $z_{\rm abs}= 3.077589$  & $z_{\rm abs}= 3.076653$  & $z_{\rm abs}= 3.076331$  & & &\\ \hline \hline
H\,{\sc i}   & 1215             & 12.00 & --- & --- & --- & 20.32 $\pm$ 0.05   &  ---             &  ---   \\
C\,{\sc ii}  & 1036, 1334       & 8.43  & 13.07 $\pm$ 0.03 & 12.99 $\pm$ 0.04 &  12.83 $\pm$ 0.06 &
13.33 $\pm$ 0.02  & $-3.42\pm0.06$ &  $>+0.24$  \\
O\,{\sc i}   & 988, 1302        & 8.69  & 13.70 $\pm$ 0.02 & 13.55 $\pm$ 0.03 & --- & 
13.93 $\pm$ 0.02  & $-3.08\pm0.05$ &  $>+0.58$   \\
Al\,{\sc ii} & 1670             & 6.45  & $\leq10.95^{\rm a}$ & --- & --- &
$\leq10.95^{\rm a}$   & $\leq-3.82^{\rm a}$ &   ---  \\
Si\,{\sc ii} & 1260, 1526       & 7.51  & 12.40 $\pm$ 0.01 & 12.20 $\pm$ 0.03 &  12.50 $\pm$ 0.02 &
 12.61 $\pm$ 0.01 & $-3.22\pm0.05$ &  $>+0.44$  \\
Fe\,{\sc ii} & 1260, 1608, 2382 & 7.47  & $\leq11.83^{\rm a}$& $\leq11.83^{\rm a}$ & --- &
$\leq12.13^{\rm a}$  & $\leq-3.66^{\rm a}$ &  ---                 \\ 
\hline
\end{tabular}
\begin{tabular}{l}
     $^{*}$ We adopt the solar abundances reported in \citet{Asplund2009}. These are consistent with those recommended by \\ \citet{Lodders2019} when considering the associated uncertainties. \\
     $^{a}$ $2\sigma$ upper limit on column density. \\
     The total column densities and abundance ratios are based on the chemistry of the predominantly neutral components \\
     (i.e., Comp 1 and 2). \\
\end{tabular}
\end{table*}

\section{Chemical modelling}

\subsection{Chemical enrichment model}
\label{sec:model}
 Using the observed abundance pattern of this DLA, alongside the stochastic chemical enrichment model described in \citet{Welsh2019}, we investigate the possible enrichment history of this DLA. The basis of our model is described briefly below. The singular update since previous applications is the treatment of upper-limits. When accounting for the error associated with the observed data we take two approaches. For abundances with known errors, this distribution is modelled with a Gaussian distribution function. For abundances with upper limits, this distribution is modelled by a Q-function. 

\subsubsection{Likelihood modelling technique}

The mass distribution of Population III stars is modelled as a power-law: $\xi(M)=k\,M^{-\alpha}$, where $\alpha$ is the power-law slope\footnote{In our formulation, $\alpha=2.35$ corresponds to a Salpeter IMF.}, and $k$ is a multiplicative constant that is set by defining the number of stars, $N_{\star}$, that form between a minimum mass $M_{\rm min}$ and maximum mass $M_{\rm max}$, given by:
\begin{equation}
\label{eqn:N}
\centering
    N_{\star} = \int_{M_{\rm min}}^{M_{\rm max}} kM^{-\alpha} dM  \; .
\end{equation}
In this work, $N_{\star}$ represents the number of stars that have contributed to the enrichment of a system. Since the first stars are thought to form in small multiples, this underlying mass distribution is necessarily stochastically sampled. We utilise the yields from simulations of stellar evolution to construct the expected distribution of chemical abundances given an underlying IMF model. These distributions can then be used to assess the likelihood of the observed DLA abundances given an enrichment model.

We utilise the yields from \citet{HegerWoosley2010} (hereafter \citetalias{HegerWoosley2010}) to construct the expected distribution of chemical abundances given an underlying IMF model. The \citetalias{HegerWoosley2010} simulations trace nucleosynthesis processes within Population III progenitors across their lifetimes and calculate the ejected yields of elements following the eventual collapse of these stars as core-collapse SNe (CCSNe). These simulations explore initial progenitor masses in the range $M=(10-100)~{\rm M_{\odot}} $, explosion energies from $E_{\rm exp}=(0.3-10)\times10^{51}$ erg, and mixing parameters from $f_{\rm He} = 0 - 0.25$. The explosion energy is a measure of the final kinetic energy of the ejecta at infinity, while the mixing between stellar layers is parameterised as a fraction of the helium core size. This parameter space is evaluated across 120 masses, 10 explosion energies, and 14 mixing parameters. We linearly interpolate between this grid of yields during our analysis. 
We refer the reader to \citetalias{HegerWoosley2010} and \citet{Welsh2019} for further details and caveats related to these yield calculations. 

 Our model contains six parameters: $N_{\star}$, $\alpha$, $M_{\rm min}$, $M_{\rm max}$, $E_{\rm exp}$ and $f_{\rm He}$. The range of model parameters we consider are:
\begin{eqnarray*}
-5 \leq &\alpha& \leq  5 \; , \\ 
 1 \leq &N_{\star}& \leq 150 \; , \\
 0.3 \leq &E_{\rm exp}/10^{51}{\rm erg}& \leq 10  \; ,\\
 12 \leq & M_{\rm max}/ {\rm M_{\odot}}& \leq 70 \; .
\end{eqnarray*}
In what follows, we assume that stars with masses $>10~{\rm M_{\odot}}$ are physically capable of undergoing core-collapse. Therefore, we fix $M_{\rm min}=10~{\rm M_{\odot}}$. We also assume that the mixing during the explosive burning phase of the SN occurs within a region $10$~per cent of the helium core size (i.e., $f_{\rm He}=0.100$). This is the fiducial choice from \citetalias{HegerWoosley2010} based on their comparison with the \citet{Cayrel2004} stellar sample. These empirically motivated constraints reduce the number of free parameters to 4. This ensures the remaining model parameters can be optimally estimated based on the number of currently known abundance ratios. We apply uniform priors to all remaining model parameters. Note that the upper bound on $M_{\rm max}$ corresponds to the mass limit above which pulsational pair-instability SNe are believed to occur \citep{Woosley2017}. We allow $M_{\rm max}$ to be a free parameter to account for any `islands of explodability' that could influence the collapse of these progenitors; simulations from \citet{Sukhbold2018} find that there may be a mass above which the entire star collapses directly to a black hole. Given this possibility, our model utilises the assumption that stars born with a mass above $M_{\rm max}$ will collapse directly to black holes without any chemical yield.

\begin{figure*}
    \centering
    \includegraphics[width=0.9\textwidth]{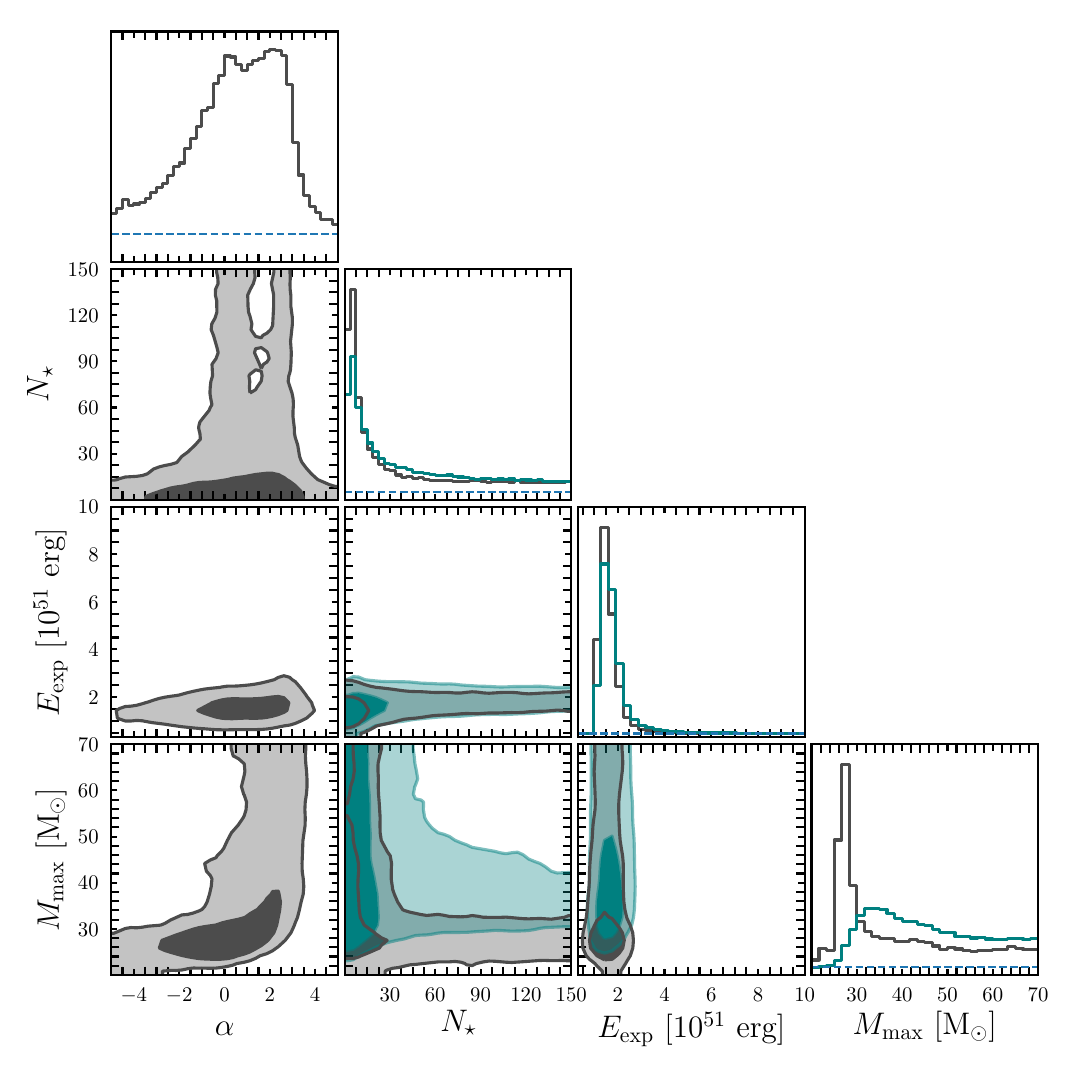}
    \caption{The marginalised maximum likelihood distributions of our fiducial model parameters (main diagonal), and their associated 2D projections, given the known chemistry of the DLA towards \jqso. The dark and light contours show the $68\%$ and $95\%$ confidence regions of these projections respectively. The horizontal blue dashed lines mark where the individual parameter likelihood distributions fall to zero. The grey distributions correspond to the analysis of the full parameter space, described in Section~\ref{sec:model}. The green distributions are the result of imposing a Salpeter slope for the IMF (i.e. $\alpha=2.35$).}
    \label{fig:j0903_MCMC}
\end{figure*}

\subsection{Application to DLA data}

\begin{figure*}
    \centering
    \includegraphics[width=\textwidth]{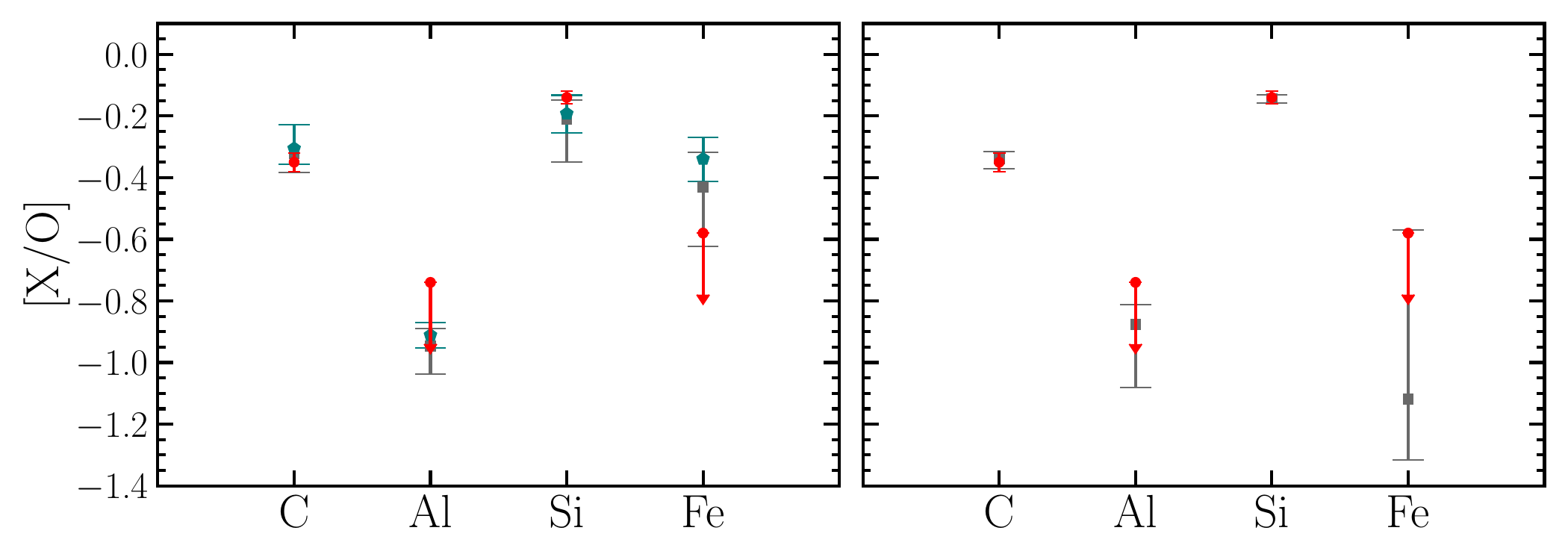}
    
    \caption{Left: Modelled abundances (grey) vs. observed data (red) given a stochastic enrichment model with a free IMF slope. The model successfully reproduces all chemical abundance measurements. The grey data show the median predicted value and the interquartile range of predicted abundances given the data. The modelled abundances given a Salpeter IMF are shown in green. Right:  Modelled abundances (grey) vs. observed data (red) under the assumption that the environment has been enriched by 1 Population III SN. }
    \label{fig:dla_dat_v_model}
\end{figure*}

Using all available chemical abundance measurements of the DLA towards J0903+2628 we run a Markov Chain Monte Carlo (MCMC) maximum likelihood analysis to find the enrichment model that best fits these data. The converged chains are shown in Figure~\ref{fig:j0903_MCMC} for both a Salpeter IMF (green histograms and contours) and a power-law IMF where the slope is a free parameter (grey histograms and contours). The inferred enrichment models are broadly consistent for both the choice of a varying and fixed IMF slope. This is to be expected since the maximum likelihood estimate of $\alpha$ for the varying IMF slope is consistent with that of a Salpeter distribution at the 95 per cent confidence level. However, we note there is preference towards a flat or even top-heavy IMF slope. The inferred number of enriching stars shows a preference towards low values of $N_{\star}$. The distribution of the typical explosion energy is more clearly centred around $E_{\exp} \sim1.8\times10^{51}$~erg. Both the distribution of $N_{\star}$ and $E_{\rm exp}$ are stable against changes to the slope of the IMF. In contrast to this, the inferred maximum mass of the enriching star is dependent on the slope of the IMF. For a Salpeter distribution, there is a preference towards $M_{\rm max} > 40~{\rm M_{\odot}}$; this can be seen from the plateau in the distribution shown by the green curve in the bottom right panel of Figure~\ref{fig:j0903_MCMC}. For the case of a varying IMF slope, there is a peak in the distribution of the maximum mass at $M_{\rm max} \sim25~{\rm M_{\odot}}$. This is driven by the possibility of both flat and top-heavy IMFs. This can be seen by the correlation between $\alpha$ and $M_{\rm max}$ in the bottom left panel of Figure~\ref{fig:j0903_MCMC}. These results are consistent with our previous investigations of the enrichment of metal-poor DLAs.

The broad distributions in Figure~\ref{fig:j0903_MCMC} highlight the lack of preference towards a particular enrichment model (e.g. both the number of enriching stars and the maximum stellar mass are relatively unconstrained). To investigate the origin of this, we test how well the inferred model can replicate the data. Using our stochastic enrichment model, we can recover the predicted abundance pattern given the inferred parameter distributions. Figure~\ref{fig:dla_dat_v_model} (left panel) shows the observed abundances (red symbols) alongside the modelled abundances for the case of a free IMF slope (grey symbols) and for a Salpeter IMF (green symbols). While the observed abundance pattern is well-recovered, there is a preference towards a top-heavy IMF and a low number of enriching stars. The predicted abundances given a Salpeter IMF do not produce the same agreement with the observed abundances of the DLA. This is primarily driven by the predicted [Fe/O] abundance given a Salpeter IMF. The remaining relative abundances are predicted equally well by both scenarios. This highlights both the importance of considering multiple relative abundances simultaneously and the sensitive nature of abundance pattern fitting. We note that, given the current associated errors, there would be little distinction between the performance of either model at the $3\sigma$ level of confidence. This is further motivation to obtain high-precision abundance determinations for these rare gaseous relics. Since these data are best modelled by a low number of enriching stars, we will now consider the possibility that this system has been enriched by a single Population III SN. This idea was previously considered by \citet{Cooke2017}, and we now reevaluate the best-fitting model parameters given the new abundance information. \\

\begin{figure*}
    \centering
    \includegraphics[width=1.3\columnwidth]{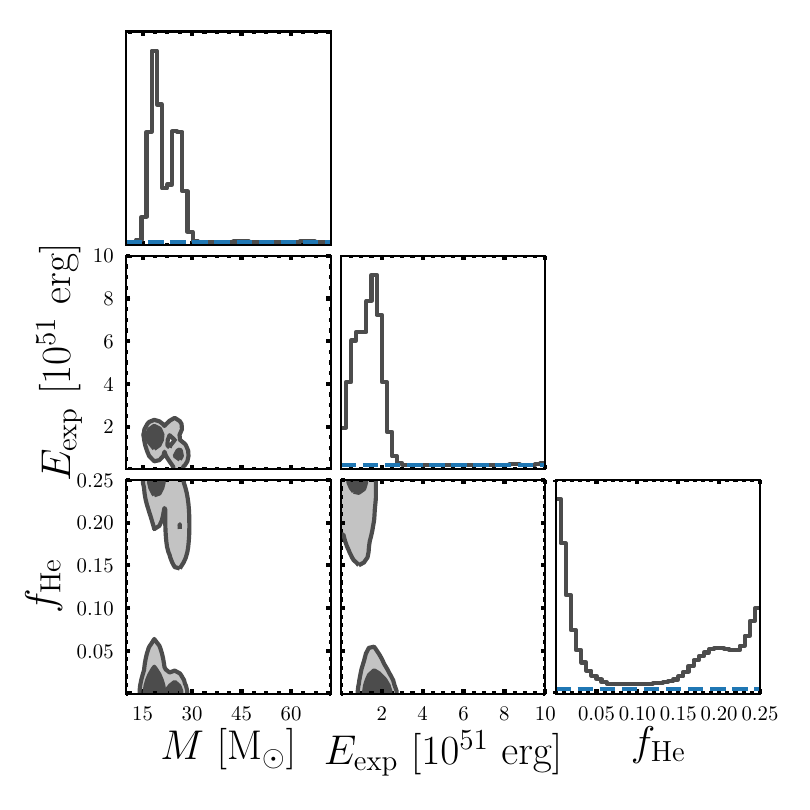}
    \caption{Results of our MCMC analysis of the chemical enrichment of the DLA towards \jqso\ assuming the number of enriching stars $N_{\star}=1$. From left to right, we show the progenitor star mass, the explosion energy, and the degree of mixing. The diagonal panels indicate the maximum likelihood posterior distributions of the model parameters while the 2D contours indicate the correlation between these parameters. The dark and light grey shaded regions indicate the 68 and 95 per cent confidence intervals, respectively. In the diagonal panels, the horizontal blue dashed line indicates the zero-level of each distribution. 
    }
    \label{fig:MCMC_1N}
\end{figure*}

\begin{figure}
    \centering
    \includegraphics[width=\columnwidth]{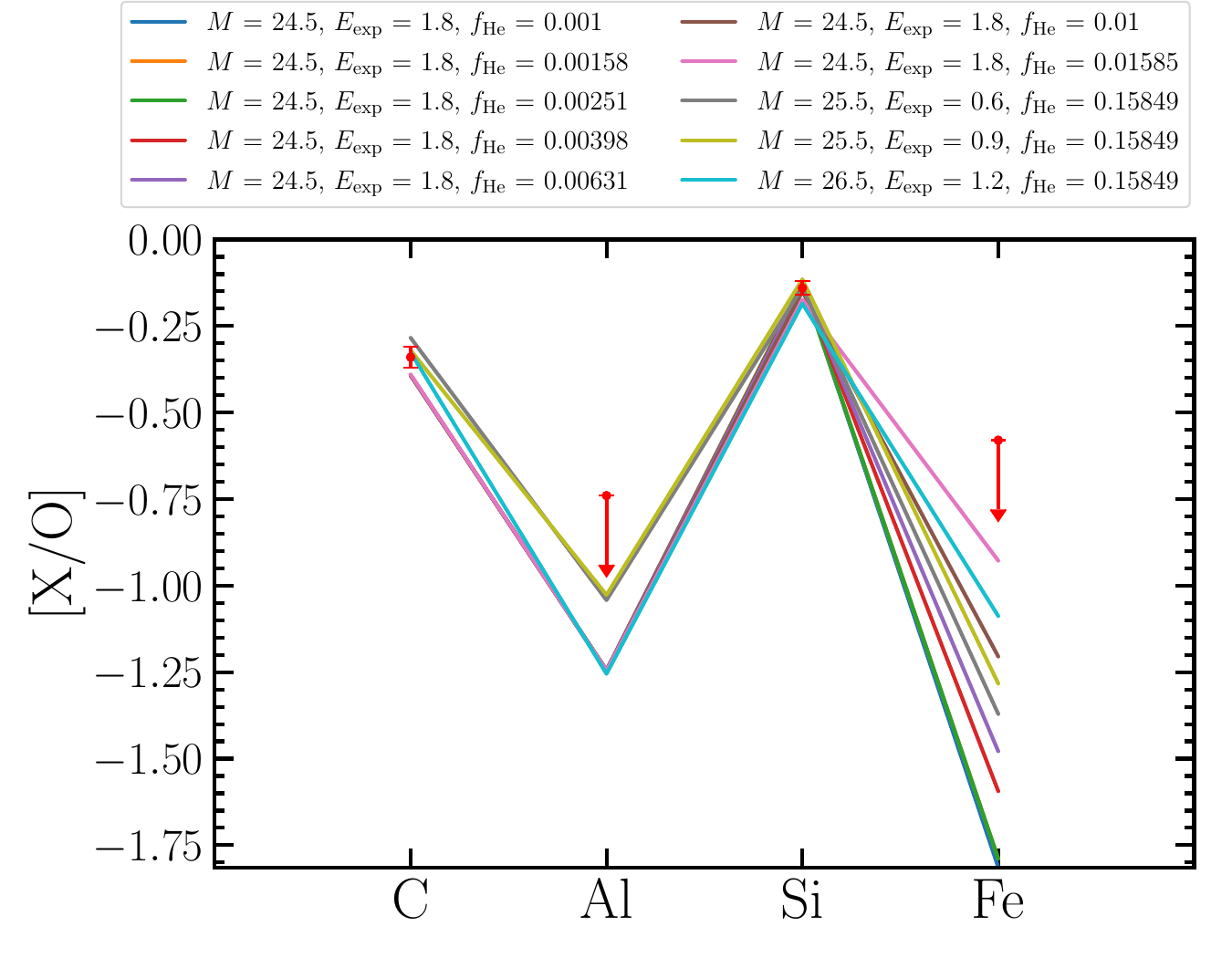}
    \caption{The predicted abundance patterns (coloured lines) for a selection of individual progenitor models that we have found are best able to model the data. The legend indicates the associated progenitor properties where $M$, $E_{\rm exp}$, and $f_{\rm He}$ are given in units of: $M_{\odot}$, $10^{51}~{\rm erg}$, and fraction of the He core size. The observed abundances are shown as red symbols. Note the similarity in model parameters barring the degree of mixing. The range of the predicted [Fe/O] abundance is responsible for the large range shown in the right panel of Figure~\ref{fig:dla_dat_v_model}. }
    \label{fig:top_models}
\end{figure}

\begin{figure*}
    \centering
    \includegraphics[width=\textwidth]{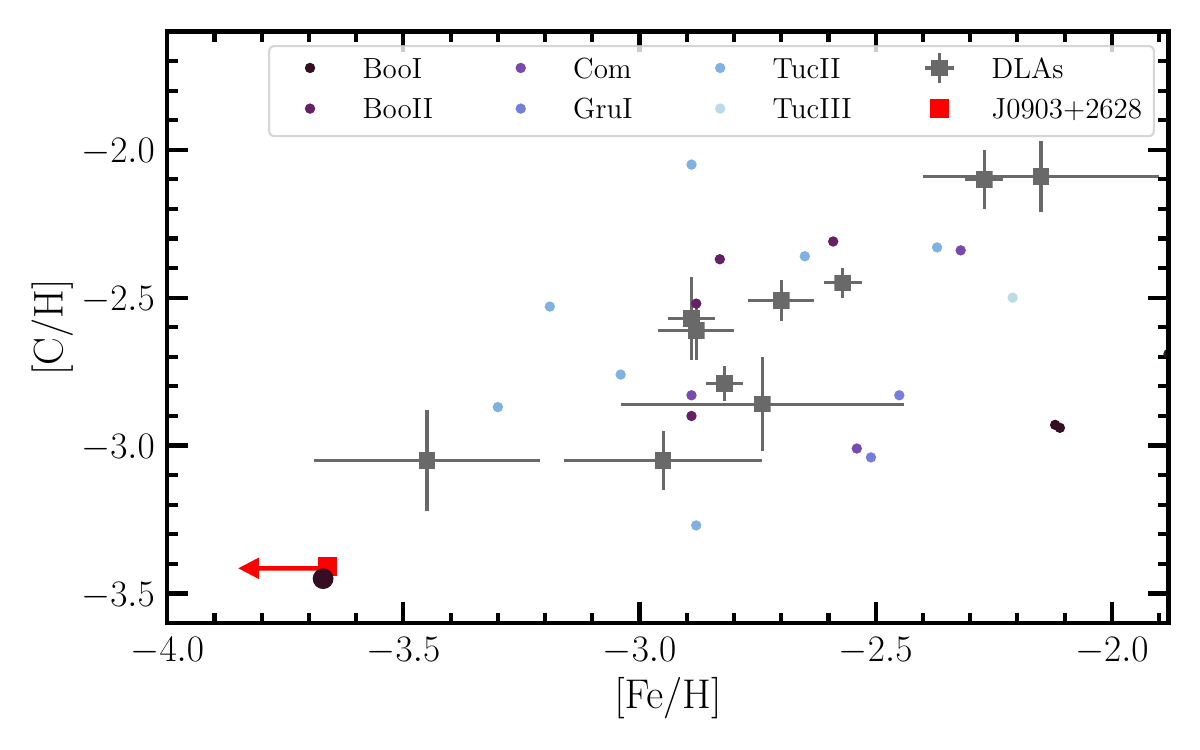}
    \caption{[C/H] vs [Fe/H] abundances of stars in ultra faint dwarf (UFD) galaxies alongside that of the most metal-poor DLAs. Stellar data are from the SAGA database \citep{Susa2014} and include the stars within these UFDs that have known [Fe/H] and [C/H] abundances with bounded errors. The stars from different systems are distinguished by their colour as described in the legend --- sequentially: Bo\"{o}tes~I, Bo\"{o}tes~II,  Coma Berenices, Grus~I,  Tucana~II, Tucana~III. The abundances of the DLA reported here are shown by the red square. For clarity, the circle of the star that occupies a similar parameter space to that of the DLA reported here (Boo$-1137$) has been enlarged. }
    
    \label{fig:stars_in_ufds}
\end{figure*}
\subsubsection{Enrichment by one Pop III SN}
The results of considering a single Pop III progenitor are shown in Figure~\ref{fig:MCMC_1N} (and the right panel of Figure~\ref{fig:dla_dat_v_model}). Note that, in this scenario, we allow the degree of mixing to vary as a free parameter with a uniform prior. 
Figure~\ref{fig:MCMC_1N} highlights that there is a bimodal distribution of the inferred initial progenitor mass, concentrated in the mass range $M=(18-27)~{\rm M_{\odot}}$. The corresponding explosion energy of this progenitor is in the range $E_{\rm exp} = (0.6-1.8)\times10^{51}~{\rm erg}~(1\sigma)$ and there is preference towards a low degree of mixing between the stellar layers, $f_{\rm He}<0.1~(1\sigma)$, though it is unconstrained at the $2\sigma$ level. The lack of information regarding the degree of mixing is curious. To investigate this further, we have explored the combination of progenitor parameters best able to individually fit the data. We isolate these models based on their log likelihood value. Figure~\ref{fig:top_models} shows a subset of these progenitor models and the associated predicted abundances. Specifically, we display the 10 progenitor models with the largest log likelihood values. We find that these models are almost identical barring the degree of mixing. Notably, these model span a large fraction of the $f_{\rm He}$ parameter space and are a driving force behind the large spread of the predicted [Fe/O] abundance shown in the right panel of Figure~\ref{fig:dla_dat_v_model}. We therefore expect that the mixing parameter can be pinned down once the Fe\,{\sc ii} column density of this system is confidently measured.  In Appendix~\ref{appen:cooke}, we compare these model distributions to those originally inferred in \citet{Cooke2017}. In summary, we find that the latest limits on [Fe/O] and [Al/O] are sufficiently informative to rule out the previously inferred enrichment scenario. 

 In Figure~\ref{fig:dla_dat_v_model} (right panel), we show the modelled abundance ratios given the distributions of progenitor parameters shown in Figure~\ref{fig:MCMC_1N}. Through the comparison of the left and right hand panels of this figure, it is clear that the scenario of enrichment by an individual Population III SN is a similarly good fit to the data as the fiducial stochastic model. The predicted abundances in both scenarios are likely being drawn from similar progenitors. The maximum likelihood parameter estimates from the stochastic model are consistent with sampling an individual $\sim25$~M$_{\odot}$ progenitor (achieved via a top-heavy IMF model with $N_{\star}=1$ and $M_{\rm max}=25~{\rm M}_{\odot}$). To further compare the relative performance of these models we consider the Akaike Information Criterion (AIC) \citep{Akaike1974}. The AIC is given by ${\rm AIC} = 2K - 2\ln (\mathcal{L})$ where $K$ is an indication of the number of free parameters\footnote{Specifically, $K = 2 + n$ where $n$ is the number of free model parameters. In the case of the stochastic model with a free IMF slope, $n=4$. For the case of an individual Pop III enricher $n=3$.} and $\mathcal{L}$ is the maximum likelihood of the model. The AIC given the stochastic model is $14$. For the individual Pop III progenitor model the $AIC=3$; this relatively lower value indicates that modelling the enrichment via one Pop III SN is better able to reproduce these data. We also note that the observed upper limit of [Fe/O] is more easily explained given the yields of an individual Population III SN. Though, the true level of agreement is reserved for definitive [Fe/O] and [Al/O] measurements. Indeed it is possible that with future data, and an increasingly extensive abundance pattern determination, our current model may be ruled out. At present, it is interesting that the known abundances of this DLA are preferably modelled by the yields of an individual Population III SN rather than a small handful of these progenitors. Furthermore, given the best fitting models (i.e. those shown in Figure~\ref{fig:top_models}), the predicted [C/Fe] abundance ratio suggest that this DLA may be in the UMP regime and could be significantly C-enhanced. 

Another possibility is that Population II stars may have contributed some of the metals to this near-pristine DLA. The abundance ratios that are often accessible for the most metal-poor DLAs are [C/O], [Si/O], and [Fe/O]. The yields of these abundance ratios are broadly similar across both Population III and Population II stars \citep[][Figure 1]{Welsh2019}. This can make it challenging to disentangle the relative contributions of these populations based on yields alone. However, the tell-tale sign of enrichment from a primordial population of stars may be seen through empirical trends or, indeed, inferences of our model parameters --- for example: a non-Salpeter IMF slope, a low number of enriching stars, or, an exotic explosion energy. Crucially, we do see this preference towards a low number of enriching stars in our analysis (Figure~\ref{fig:j0903_MCMC} - top panel, second column). Further, if this were a Population II enriched system, we may expect the observed abundances to be consistent with the IMF-weighted abundances given a Salpeter IMF. This is not the case. Particularly, the IMF-weighted [O/Fe] abundance [O/Fe]~$=+0.46$ (assuming a typical explosion energy and mixing; $E_{\rm exp}=1.2\times10^{51}$~erg and $f_{\rm He} =0.1$) is inconsistent with the observed lower limit provided by our new data. For completeness, we have also repeated our analysis under the assumption of enrichment by an individual Population II star and found that, using the yields from \citet{WoosleyWeaver1995}, the enrichment by a Population III star still provides the best fit to the data given the currently available yields (see Appendix~\ref{appen:popii} for further details). The case of enrichment by an individual Population III SN is often reserved for the stellar relics found in the halo of the Milky Way and surrounding dwarf galaxies. In the following section, we discuss the implications of these results and draw comparisons with the chemistry of other relic objects. Ultimately, the goal is to reveal the nature of this most metal-poor DLA and its place within galaxy evolution.

\section{Discussion}
\label{sec:disc}
The nature of DLAs, especially the most metal-poor systems, is still an open question. Various interpretations have arisen since the first DLA survey. These highest H\,{\sc i} column density absorption line systems could be associated with the discs of high redshift galaxies or, indeed, a wider range of galaxy masses and morphologies \citep{wolfe1986, Pettini1990, Haehnelt1998}. The latter scenario is supported by the results of both modern hydrodynamical cosmological simulations \citep[e.g.][]{Pontzen2008, Bird2013, Rahmati2014}, and dedicated imaging surveys \citep[e.g.][]{Fumagalli2015}. Searches for the emission from host galaxies aim to illuminate this issue. The use of integral field spectrographs have proven to be fruitful tools for these searches \citep[e.g.][]{Peroux2011, Peroux2012, Fumagalli2017, Mackenzie2019, Berg2022, Lofthouse2022}; as has the Atacama Large Millimeter/submillimeter Array (ALMA) \citep[e.g.][]{Moller2018, Neeleman2018, Klitsch2021}. Surveys with these instruments may tentatively indicate that there is a more complicated relationship between the column density of the absorbing gas and its association with the host galaxy than initially thought \citep[e.g.][]{Lofthouse2022}.

Specifically considering the environments of DLAs, the detection rate of associated galaxies has increased in recent years \citep[e.g.][]{Mackenzie2019, Klitsch2021}. However, the presence of multiple galaxies (alongside their typically large impact parameters) in the observed fields make it challenging to assign a single host galaxy to the DLA. There are further complications related to the fraction of the DLA population associated with active star formation. It has been suggested that there may be a steep relationship between the emission luminosity of the host galaxy and the metallicity of the DLA \citep{Krogager2017}. The true nature of this relationship is an open question. If a mass-metallicity relation applies to galaxies at $z=2-3$, the most metal-poor DLAs are likely to be associated with the faintest (and therefore most difficult to detect directly) host galaxies. Moreover, there is yet to be a detection in emission of a host galaxy associated with one of the four known EMP DLAs. 
Consequently, we look to other means to understand the association between the most metal-poor DLAs and galaxy evolution. 

 The chemistry of these environments is an invaluable tool to link these gaseous relics seen at high redshift to their local descendants. This is a necessary step to understand the role of near-pristine DLAs in galaxy evolution. Furthermore, it facilitates comparisons with other relic environments that may reveal similar enrichment histories and shared evolutionary pathways. For this reason, the following section is dedicated to comparing the chemistry of the most metal-poor DLA known to the stars within ultra-faint dwarf galaxies (UFDs). 
\subsection{Comparison with UFDs}
We now compare the chemistry of this DLA to the chemistry of stars in the ultra faint dwarf galaxies \citep[UFDs; $M_{\rm V} >-7.7$ from the definition by][]{Simon2019} that orbit the Milky Way. The lowest mean metallicity of a confirmed UFD is that of Tucana~II with [Fe/H]~$=-2.9^{+0.15}_{-0.16}$ \citep{Bechtol2015, Walker2016, Chiti2018, Simon2019}. These ancient galaxies are on the borderline of extreme metal-paucity and are considered relics of the early Universe \citep{Salvadori2009, Bovill2009, Munoz2009, BrommYoshida2011}. 

 To fairly compare the metallicity of the DLA towards \jqso\ with similar stellar relics would require the detection of the same elements in each system. Despite the wide range of element abundances available for stellar relics, there is relatively little chemical abundance overlap for the most metal-poor DLAs and stars. For the purpose of this work, we therefore reserve our analysis to comparing the relative abundance of a light atomic number element (in this case [C/H])and an Fe-peak element, [Fe/H]. Figure~\ref{fig:stars_in_ufds} shows [C/H] as a function of [Fe/H] for both metal-poor DLAs (grey squares) and stars within UFDs (color-coded circles) as reported by the SAGA database \citep{Susa2014}. We restrict our comparison to stars within UFDs with bounded [C/H] and [Fe/H] abundances. The EMP DLA reported here is shown as the red square. From this plot, we can see that the Fe-metallicity evolution of [C/H] reported for the most metal-poor DLAs is consistent with that of the stars within UFDs. We note that this may be influenced by the lack of known carbon-enhanced DLAs as well as potential observational biases due to the challenges of chemically characterising the stars within the UFDs. The faint nature of these UFDs mean that we may only be observing, and subsequently characterising, the brightest stars within these systems. This limitation will be drastically improved with the upcoming 4DWARFS survey \citep{Skuladottir2023}.

 Interestingly, there is an EMP star within Bo\"{o}tes~I (large purple circle in Figure~\ref{fig:stars_in_ufds}) that occupies a similar parameter space as the DLA reported here. 
This star is known as Boo$-1137$. The relative abundances of mutual elements (i.e., those  also detected towards \jqso) are provided in Table~\ref{tab:boo}. These are taken from \citet{Norris2010} who used the solar scale $\log~\epsilon_{Fe}=7.45$ \citep{Asplund2005}\footnote{$\log~\epsilon_{X}\equiv\log({\rm X/H})+12$. Note, we adopt the solar scale referenced in Table~\ref{tab:j0903}. In more recent stellar studies it is typical to assume $\log~\epsilon_{Fe}=7.51$. This, like our solar scale, is drawn from \citet{Asplund2009}. Though, we take the average solar Fe abundance reported from both photospheric and meteoritic determinations. On the solar scale typically used in stellar studies, our Fe limit would be [Fe/H]~$<-3.70~(2\sigma)$.}.
\begin{table}
	\centering
	\caption{Elemental abundance ratios of Boo$-1137$ from \citet{Norris2010}, corrected to the solar abundance scale adopted in our work. The quoted errors are the $1\sigma$ confidence limits while the upper limits are $2\sigma$. }
	\label{tab:boo}
	\tabcolsep=0.10cm
\begin{tabular}{lcc}
\hline
El    & [X/H]$_{\rm DLA}$  & [X/H]$_{\star}$  \\ 
 \hline \hline
C  & $-3.42\pm0.06$ & $-3.45\pm0.20$  \\
O  & $-3.08\pm0.05$ & $<-1.75$  \\ 
Al & $\leq -3.82$ & $-4.35\pm0.11$  \\ 
Si & $-3.22\pm0.05$ & $-2.89\pm 0.16^{a}$  \\ 
Fe & $\leq-3.66$ & $-3.68\pm0.11$ \\
\hline
\end{tabular}
\begin{tabular}{l}
     $^{a}$ This error given by that of [Si/Fe] which is the ratio used in \\ the MCMC enrichment model analysis.
\end{tabular}
\end{table}

\begin{figure}
    \centering
    \includegraphics[width=\columnwidth]{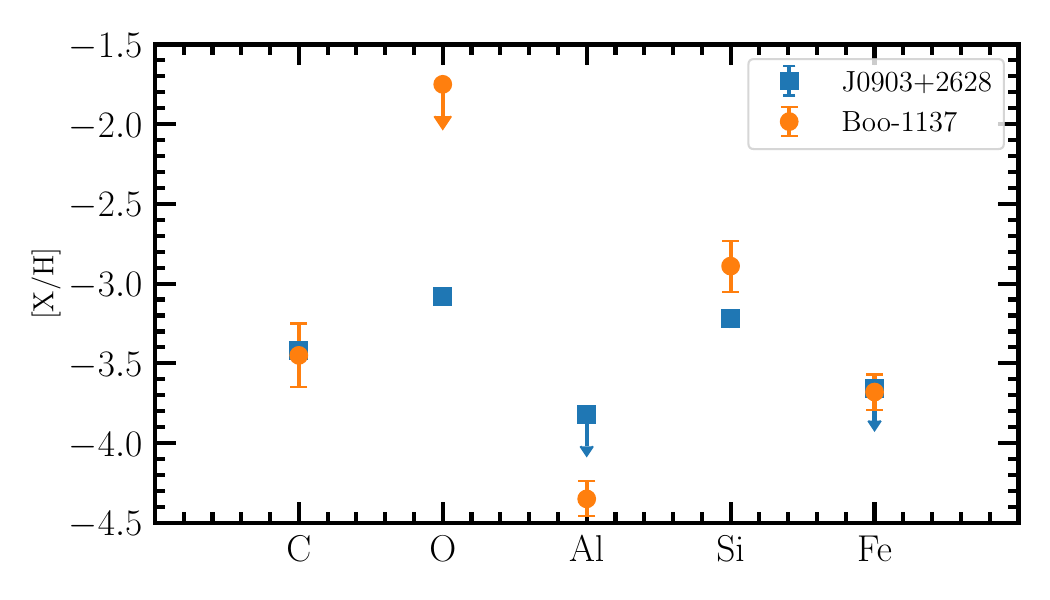}
    \caption{Comparisons of the [X/H] abundances of the chemical elements detected for both \jqso\, (blue) and Boo$-$1137 (orange). 
    }
    \label{fig:dla_vs_boo}
\end{figure}
After rescaling the [Fe/H] abundance to the solar values adopted in this paper (see Table~\ref{tab:j0903}), Figure~\ref{fig:dla_vs_boo} visualises this comparison of similar relative abundances. While the upper limits provide limited information, the chemical make-ups of these two environments (the DLA and the star Boo$-$1137) appear to be consistent with each other. We speculate that this common chemical abundance pattern may indicate a common formation channel. In other words, the similar chemistry of these environments may imply they were enriched by a similar (possibly primordial) population of stars. We note that the [Mg/C], [Fe/H], and A(C) abundances of Boo$-1137$ are indicative of enrichment via Pop III SNe given the recent analysis by \citet{Rossi2023}. We also point out that our comparison shows the chemical abundances relative to hydrogen. Consequently, the absolute abundances presented in Figure~\ref{fig:dla_vs_boo} are dependent on the amount of hydrogen that the metals have mixed with. Under the assumption that at least \emph{some} EMP stars form from EMP DLA gas, comparing the absolute abundances allows us to test how effectively the metals are mixed in EMP DLAs. The similar trends of EMP DLAs and stars seen in Figure~\ref{fig:dla_vs_boo} supports our assumption, and the slight excess dispersion seen for the stellar sample could reflect intrinsic metallicity variations within EMP DLAs (i.e. EMP DLA gas is not entirely well-mixed) that may reflect stochastic sampling of the IMF. Another possibility is that a single EMP DLA may comprise the gas and stars from chemically distinct minihaloes \citep[e.g.][]{Welsh2021}.

\begin{figure}
    \centering
    \includegraphics[width=\columnwidth]{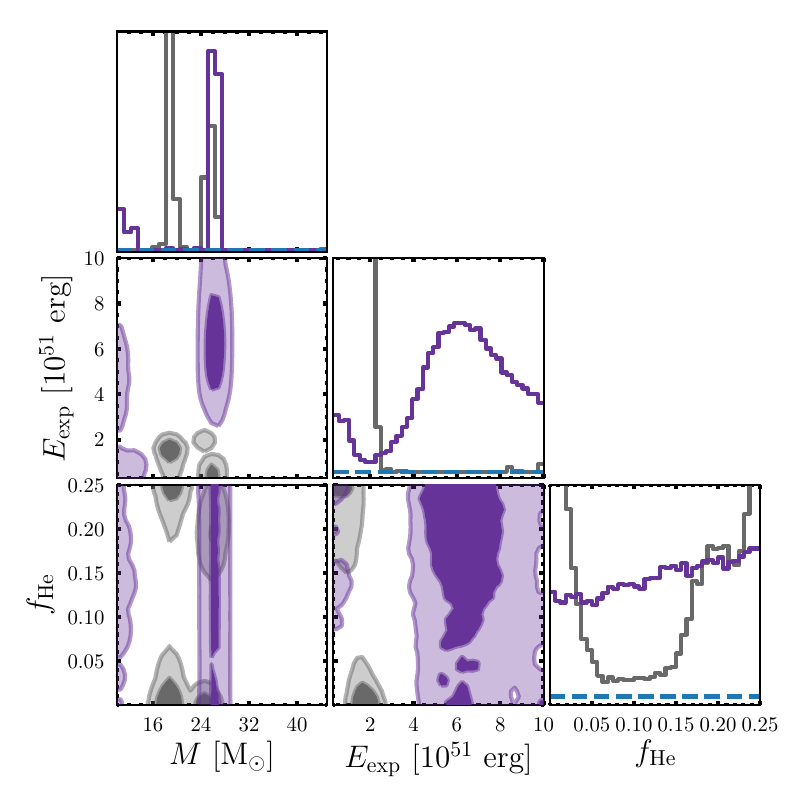}
    \caption{Same as Figure~\ref{fig:MCMC_1N}, but based on the [C/Fe], [Si/Fe], [O/Fe], and [Al/Fe] of Boo$-1137$. The distributions associated with Boo$-1137$ are shown in purple while those for \jqso\ (from Figure~\ref{fig:MCMC_1N}) are shown in grey for reference.
    }
    \label{fig:MCMC_1N_boo}
\end{figure}

\begin{figure}
    \centering
    \includegraphics[width=\columnwidth]{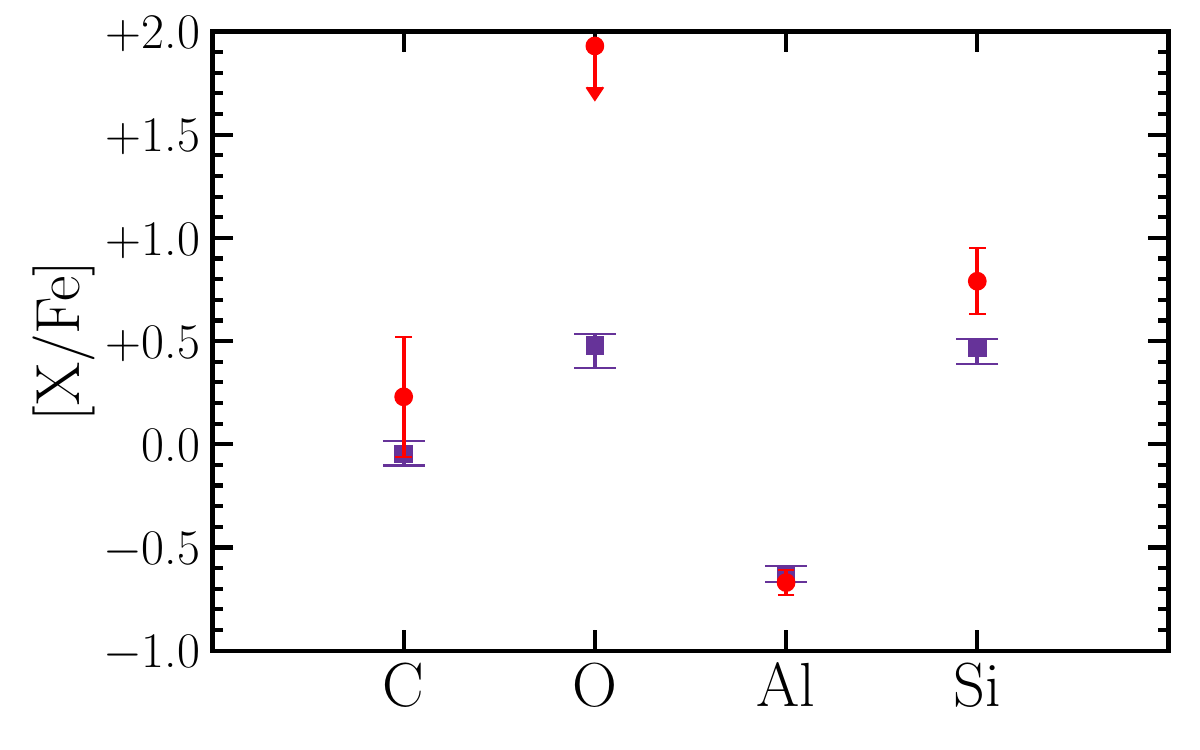}
    \caption{Same as Figure~\ref{fig:dla_dat_v_model}, but based on the enrichment model inferred for Boo$-1137$ and colours match those in Figure~\ref{fig:MCMC_1N_boo} (i.e. model results are shown in purple). 
    }
    \label{fig:1N_mvd_boo}
\end{figure}

In order to understand the properties of the enriching population, we need to consider the relative abundances of \emph{metals} (e.g. [C/Si] which is consistent for these objects). Repeating our MCMC analysis, under the assumption of enrichment from 1 Population III SN, the model for Boo$-1137$ converges on a progenitor mass that is remarkably similar to that found for \jqso. During this analysis, we utilised the same chemical elements used for \jqso. However, to accommodate the upper limit on [O/H], we consider the abundance ratios relative to Fe instead of O. The results of this analysis and the corresponding model fit to the data are shown in Figures~\ref{fig:MCMC_1N_boo} and \ref{fig:1N_mvd_boo} respectively.  

 Despite the broadly similar abundances and inferred progenitor mass, the inferred explosion energies differ; the stellar progenitor is well-modelled by a hypernova for Boo$-$1137 (i.e. $>5\times10^{51}$~erg) while the DLA progenitor is well-modelled by a more typical ($<3\times10^{51}$~erg) CCSN. A more striking agreement between the inferred progenitor properties is needed to provide a compelling argument for a singular enrichment channel. We are currently limited by the (lack of) information provided by the observed upper limits. Indeed, the [O/Fe] abundance limit of Boo$-$1137 is relatively weak ([O/Fe]$_{\star}$~$<+1.9$) while that of \jqso\ is not ([O/Fe]$_{\rm DLA}$~$>+0.58$). This is an important informant of the explosion energy  \citep[][see also Appendix~\ref{appen:cooke}]{HegerWoosley2010, ChieffiLimongi2018}. Furthermore, we could improve our determination of the explosion energy by analysing the relative abundances of the Fe-peak elements of this DLA \citep{Cooke2013}. Using the AIC, we have checked which model is best able to reproduce the respective data of these objects. The model associated with the DLA data has a relatively lower AIC value (recall AIC~$=3$) than the model associated with the stellar data (AIC~$=5$). The relatively weaker performance of the Boo-$1137$ model can be seen in Figure~\ref{fig:1N_mvd_boo}. This figure shows that the model is not able to reproduce the observed [Si/Fe] abundance within the 1$\sigma$ errors when simultaneously reproducing the other abundance ratios.

 As things stand, we need more data to link the most metal-poor DLA currently known to local analogues. This information can be gained in a multitude of ways. First, deeper NIR observations from the ground would allow the determination of the Fe column density through the \fe{2344} and \fe{2382} features (rather than the upper limit reported here). Furthermore, within the next decade, we will see the advent of the next generation of $30-40$~m optical/IR ground-based telescopes, including the Extremely Large Telescope \citep[ELT;][]{eltpaper}, the Giant Magellan Telescope \citep[GMT;][]{gmtpaper}, and the Thirty Meter Telescope \citep[TMT;][]{tmtpaper}. With these facilities we will be able to study the \emph{most} metal-poor DLAs with unprecedented sensitivity. Thus, we will be able to determine the abundances of elements whose features are too weak to efficiently observe with current facilities (e.g. the N, Mg, Al, S, Cr, Fe, Ni, and Zn abundances of \jqso). 
 Beyond chemistry, we can use additional diagnostics (e.g. kinematics, density models, host galaxy emission, environment etc.) to investigate the properties of these ancient relics and their surrounding galaxy populations.

\subsection{Observations with JWST}
With the James Webb Space Telescope (JWST), it may be possible to detect light directly from the first stars at high redshift. However, depending on the adopted formation formalism, it may not be feasible to detect light from the first stellar population at $z>8$ without the aide of lensing \citep{Bovill2022}; see also \citet{Trussler2022} for a more optimistic outlook. 

 The most metal-poor DLAs, and indeed near-pristine systems with lower column densities, provide a signpost to some of the least chemically polluted environments at $z_{\rm abs}<6$. Such environments might be ideal to conduct searches for Population III stars at lower, more accessible, redshifts. Depending on the redshift of the absorber (e.g. $3<z_{\rm abs}<6$), the strong He\,{\sc ii} $\lambda1640$~\AA\ feature expected from Population III stars may be most efficiently studied from the ground or from space (e.g. with HST/WFC3 or JWST/NIRCam). Once the strong He\,{\sc ii} $\lambda1640$~\AA\ emission has been identified, JWST/NIRSpec is the optimal instrument to confirm the Population III nature of this emission by searching for strong H Balmer emission lines (indicative of an H\,{\sc ii} region), simultaneously combined with the absence of associated metal emission (e.g. [O\,{\sc iii}] $\lambda$ 5007~\AA). \\
\indent We have performed an initial search for exotic stellar emission using the spectroscopic data presented here. Specifically, we search for any strong He\,{\sc ii} $\lambda1640$~\AA\ emission at the redshift of the absorber (luminosity distance D$_{L}=26.8$~Gpc) and do not detect anything. We model both the continuum and the potential feature using {\sc alis}. The resulting $3\sigma$ upper limit on the flux density $f_{\lambda}<8.5\times10^{-17}$~erg~cm$^{-2}$~s$^{-1}$~\AA$^{-1}$ can be used to calculate the limit on the He\,{\sc ii} line luminosity
after assuming an emission line width consistent with that found for local ($z<0.01$) EMP galaxies \citep[FWHM$_{\rm He} = 0.6$~\AA;][]{Senchyna2019} and adopting the \citet{Planck2018} cosmology ($H_{0}=67.4\pm0.5~{\rm kms^{-1}Mpc^{-1}}$ and $\Omega_{m}=0.315\pm0.007$). 
We place a conservative upper limit on the  He\,{\sc ii} line luminosity of $<4.3\times10^{42}$~erg/s ($3\sigma)$. This is over an order of magnitude above the predicted line luminosities from \citet{Schaerer2003}. Note that, this upper limit is only applicable to possible emission that is spatially coincident with the observed DLA. Without this proximity, we do not expect the stellar light to be covered by the slit used during these observations. To evaluate the possible star formation in the vicinity of this DLA, we would benefit from additional information beyond an individual sightline. \\ 
 \indent Given the challenges associated with detecting Population III stars at $z>8$, the most metal-poor DLAs may be invaluable tools in the search for light from the first stars. Over the coming years, observational surveys are necessary to explore how efficient these DLAs are as beacons for the first stars and galaxies, and more generally, to study the most metal barren environments in the Universe. Furthermore, the firm detection of a Population III stellar signature associated with these DLA host galaxies would confirm the relic nature of these environments.

\section{Conclusions}
\label{sec:conc}
We present the updated chemical abundances of the DLA towards \jqso\, based on data collected with Keck I/HIRES. These data reaffirm that this gas reservoir is the most metal-poor DLA currently known.  Our main conclusions are as follows:

\begin{enumerate}
    \item The latest data provide an upper limit on the iron abundance, [Fe/H]$\leq -3.66~(2\sigma)$. This is a 0.85~dex improvement on the previous upper limit and confirms that this DLA is indeed EMP, and could possibly be the first UMP DLA. To secure the first firm detection of Fe in this system will require further observations. This would require a significant time investment on $8-10$~m class telescopes. Alternatively, the required sensitivity will be efficiently achieved with the next generation of $30-40$~m telescopes.
    
    \item The current data indicate that \emph{all} of the detected  $\alpha-$elements (C, O, and Si) are enhanced relative to iron, compared to a typical very metal-poor DLA. We speculate that this intriguing gas cloud may have an iron abundance considerably below the current detection limit. Such an Fe-deficient abundance pattern is reminiscent of the $\alpha$-enhanced (or, rather, Fe-deficient) stars in the halo of the Milky Way.

        \item We apply a stochastic chemical enrichment model to investigate the properties of the stars that have enriched this gaseous relic. The results of this analysis suggest that the chemistry of this environment is best modelled by a low number of enriching stars. The observed abundance pattern is well-modelled by an individual Population III SN with an initial progenitor mass $M=(18-27){\rm M_{\odot}}~(2\sigma)$. 
        
    \item We compare the chemistry of this DLA to the abundances of the most metal-poor stars in the UFDs orbiting the Milky Way. The star closest in chemistry to that of the DLA towards \jqso\ is Boo$-$1137 in Bo\"{o}tes~I with [Fe/H]~$=-3.68\pm0.11$ \citep[registered onto our adopted solar scale;][]{Norris2010}. The [X/H] abundances that are measured in both relics are broadly consistent (though we are often restricted to upper limits). This may indicate that Boo$-$1137 formed from gas that was enriched through a similar mechanism to that of \jqso. Indeed, our analysis tentatively supports the scenario in which these relics have been enriched by a single Population III progenitor with similar mass properties but with notably distinct explosion properties.
 \end{enumerate}

Our work highlights that, while rare, near-pristine DLAs are an invaluable probe of the metals produced by some of the earliest stars in the Universe. These gaseous relics may also provide a signpost to some of the least polluted environments of the Universe and thus act as beacons to search for light from the first stars directly using JWST. 

\begin{acknowledgments}
We thank the referee for a constructive report that improved the presentation of this work. We thank Rajeshwari Dutta and both the organisers and participants of the first stars IFPU focus week (15$–$19 May 2023) for helpful discussions. This paper is based on observations collected at the W. M. Keck Observatory which is operated as a scientific partnership among the California Institute of Technology, the University of California and the National Aeronautics and Space Administration. The Observatory was made possible by the generous financial support of the W. M. Keck Foundation. The authors wish to recognize and acknowledge the very significant cultural role and reverence that the summit of Maunakea has always had within the indigenous Hawaiian community. We are most fortunate to have the opportunity to conduct observations from this mountain. We are also grateful to the staff astronomers at Keck Observatory for their assistance with the observations. 
\\\\
This work has been supported by Fondazione Cariplo, grant No 2018-2329. 
During this work, R.~J.~C. was supported by a
Royal Society University Research Fellowship.
We acknowledge support from STFC (ST/T000244/1).
This project has received funding from the European Research Council (ERC) under the European Union's Horizon 2020 research and innovation programme (grant agreement No 757535). 
This work used the DiRAC@Durham facility managed by the Institute for Computational Cosmology on behalf of the STFC DiRAC HPC Facility (www.dirac.ac.uk). The equipment was funded by BEIS capital funding via STFC capital grants ST/K00042X/1, ST/P002293/1, ST/R002371/1 and ST/S002502/1, Durham University and STFC operations grant ST/R000832/1. DiRAC is part of the National e-Infrastructure. 
This research has made use of NASA's Astrophysics Data System.
\end{acknowledgments}

\section*{Data Availability}
The data underlying this article are available through the Keck Observatory Archive. The processed data used by the authors is available via request.

\facilities{Keck:I (HIRES)}

%

\vspace{5mm}


\software{
\noindent Astropy \citep{ASTROPY}, 
Corner \citep{CORNER},
EMCEE \citep{EMCEE}, 
Matplotlib \citep{MATPLOTLIB},
and NumPy \citep{NUMPY}. }


\appendix
\section{Comparison with Cooke et al. (2017)}
\label{appen:cooke}
\begin{figure*}
    \centering
    \includegraphics[width=0.8\columnwidth]{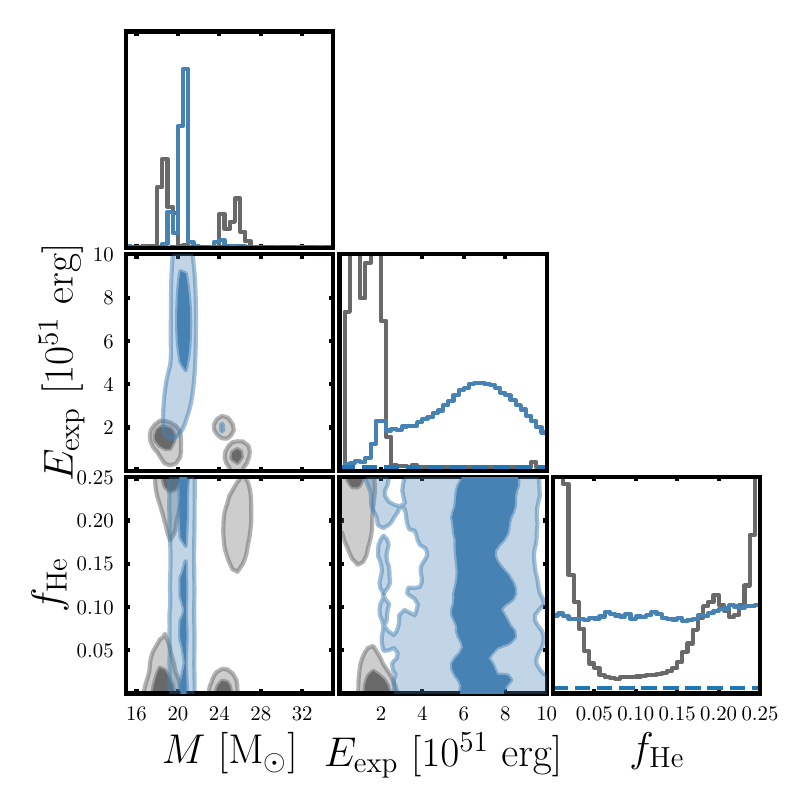}
    \caption{Same as Figure~\ref{fig:MCMC_1N} (grey contours) overplotted with the parameter distributions presented in \citetalias{Cooke2017} (blue contours).
    }
    \label{fig:MCMC_1N_cooke}
\end{figure*}

In this appendix, we compare the results of our MCMC model analysis to that determined using the original data reported in \citet{Cooke2017} (hereafter \citetalias{Cooke2017}). Specifically, we compare the inferred properties of an individual Pop III progenitor that may have enriched this gas cloud. The data presented in \citetalias{Cooke2017} facilitated the high-precision determination of both the [C/O] and [Si/O] abundances of \jqso: [C/O]~$=-0.38 \pm 0.03$ and [Si/O]~$=-0.16 \pm 0.02$. These data also provided the first limit on the [Fe/O] abundance ratio: [Fe/O]~$<+0.23$. Given that the typical [Fe/O] abundance of very metal-poor DLAs (i.e. $-3<$~[Fe/H]~$<-2$) is [Fe/O]~$\sim-0.40$, this limit is not particularly informative \citep{Cooke2011a, Welsh2022}. As discussed in the introduction, the original analysis was constrained by the original wavelength coverage of the data. The data presented in this work allow us to improve the constraint on [O/Fe] by $\sim 1$~dex and further provide the abundance limit of an additional element [Al/O]. As such, it is illuminating to explore how the information contained in these additional abundance ratio determinations inform our inferred enrichment model parameters; this comparison is shown in Figure~\ref{fig:MCMC_1N_cooke}. The grey contours are identical to those presented in Figure~\ref{fig:MCMC_1N} while the blue contours correspond to those presented in Figure 3 of \citetalias{Cooke2017}. Note that the MCMC modelling procedure adopted for both models is consistent. Explicitly, the abundance ratios used to inform the blue contours are [C/O] and [Si/O]. These are also used to inform the grey contours with the additional lower limits placed on [Fe/O] and [Al/O].

\begin{figure*}
        \centering
    \includegraphics[width=0.5\columnwidth]{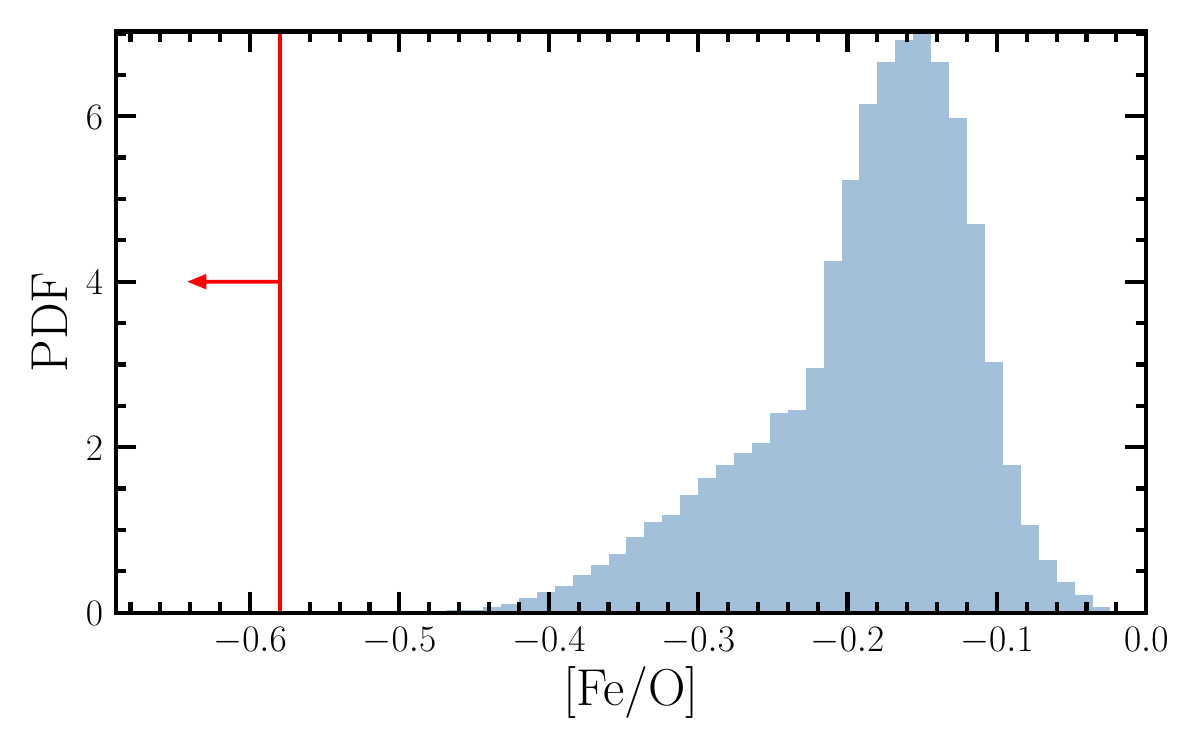}
    \caption{The predicted distribution of [Fe/O] (blue histogram) based on the model parameters inferred from the \citetalias{Cooke2017} analysis of \jqso\ compared to the limit given the latest observational data (red line and arrow).
    }
    \label{fig:feo_cooke}
\end{figure*}

Analysing the full complement of data, alongside a technique that utilises abundance ratios with lower limits, we infer different model parameters to those of \citetalias{Cooke2017}. The inferred progenitor masses are not too dissimilar. The most notable distinction is the preference towards low explosion energies when using the latest data and the preference towards a hypernova when exclusively considering the \citetalias{Cooke2017} data. Figure~\ref{fig:feo_cooke} may explain a driving force behind this difference. This figure showcases the expected distribution of [Fe/O] based on the Pop III progenitor properties inferred by \citetalias{Cooke2017} alongside the limit based on the observational data reported here. From this, it is clear that the observed limit on [Fe/O] is not reproducible given the \citetalias{Cooke2017} progenitor properties. As mentioned in Section~\ref{sec:disc}, [Fe/O] is an important informant of an enriching progenitor's explosion properties. Indeed, this can be seen by the marked improvement on the constraints of $E_{\rm exp}$. 

Ultimately, this comparison highlights that these new data provide critical information on the heavy elements and odd-atomic number elements present in this gas cloud that allow us to better study the potential enrichment history of this most metal-poor DLA. The latest model represents the progenitor properties best able to reproduce the current data. The power of our enrichment model analysis comes from simultaneously reproducing the observed abundances of relic objects. As this comparison highlights, these progenitor properties are influenced by the available abundance ratios and much can be inferred from a few key additional elements. As such, it is exciting to think what will be achieved with the next generation of ground-based telescopes (e.g. ELT/ANDES) and the detection of elements like Cr, Ni, and Zn.

\section{Enrichment by Population II stars}
\label{appen:popii}
\begin{figure}
    \centering
    \includegraphics[width=0.5\columnwidth]{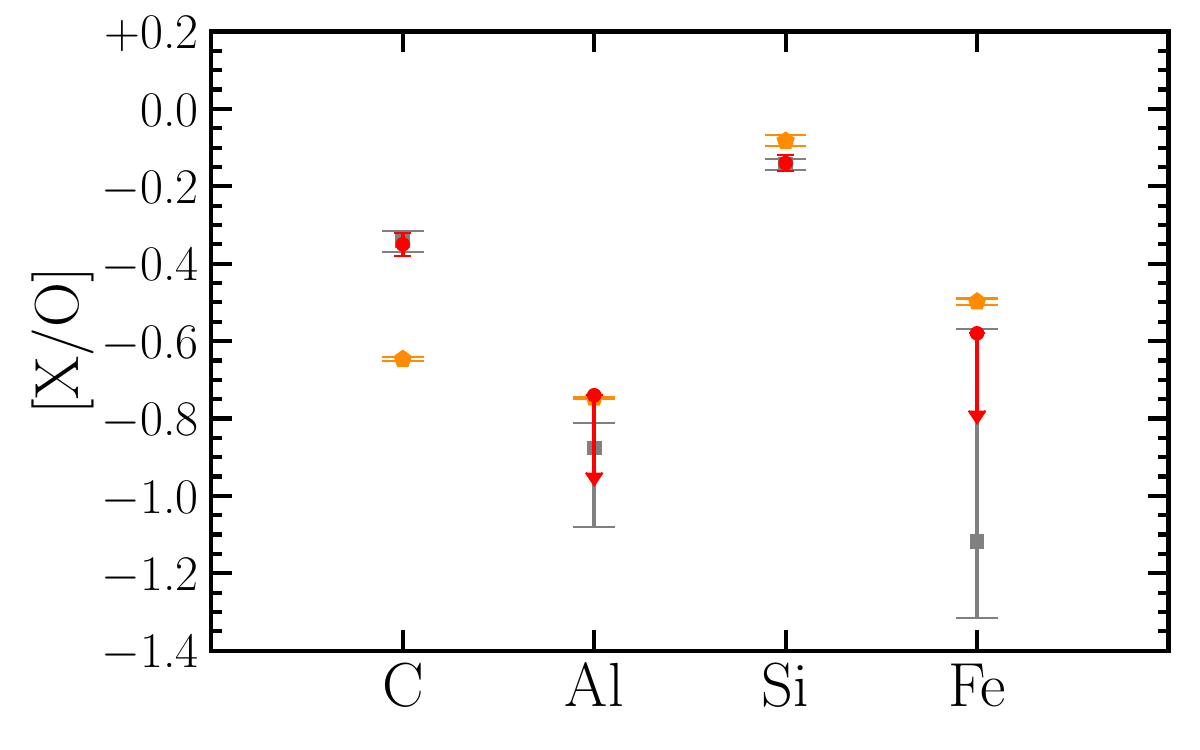}
    \caption{Same as Figure~\ref{fig:dla_dat_v_model}, but shows the enrichment model inferred for a Population II progenitor (orange) alongside those inferred for a Population III progenitor (grey). The red symbols indicate the current observationally measured abundances.
    }
    \label{fig:1N_mvd_popii}
\end{figure}

In this appendix we repeat our MCMC analysis to find the properties of 1 Population II SN that best match the abundances of the DLA reported here. For this analysis we utilise a subset of the yields from \citet{WoosleyWeaver1995} whose progenitor metallicities are $Z=0.0001Z_{\odot}$. These yields span a mass range of $M=(12-40)~{\rm M_{\odot}}$; this is the same region explored during our analysis. The yields are calculated using a fixed explosion energy and mixing parameter; therefore the only free parameter in this analysis is the progenitor mass. We find the yields are best modelled by a star will a mass of $M\sim27~{\rm M_{\odot}}$. The relative abundances estimated given the inferred Population II properties (orange symbols) are shown in Figure~\ref{fig:1N_mvd_popii} alongside the observed abundances (red symbols) and those calculated from our analysis using Population III yields (grey symbols). From this figure, we can see that the Population III yields are a better fit to the data than that provided by the Population II yields. A caveat to note is that the Population II yields are comparatively limited given the fixed explosion energy and mixing parameter adopted during the \citet{WoosleyWeaver1995} simulations. It would be interesting to see whether Population II yields that explore a similar parameter space as the \citetalias{HegerWoosley2010} yields could produce a better fitting model. However, given the publicly available yields, we conclude that a Population III progenitor is best able to replicate the observed relative abundances.

\newpage
\bibliography{references}{}

\begin{thebibliography}{}
\expandafter\ifx\csname natexlab\endcsname\relax\def\natexlab#1{#1}\fi
\providecommand{\url}[1]{\href{#1}{#1}}
\providecommand{\dodoi}[1]{doi:~\href{http://doi.org/#1}{\nolinkurl{#1}}}
\providecommand{\doeprint}[1]{\href{http://ascl.net/#1}{\nolinkurl{http://ascl.net/#1}}}
\providecommand{\doarXiv}[1]{\href{https://arxiv.org/abs/#1}{\nolinkurl{https://arxiv.org/abs/#1}}}

\bibitem[{{Abel} {et~al.}(2002){Abel}, {Bryan}, \& {Norman}}]{Abel2002}
{Abel}, T., {Bryan}, G.~L., \& {Norman}, M.~L. 2002, Science, 295, 93,
  \dodoi{10.1126/science.295.5552.93}

\bibitem[{{Akaike}(1974)}]{Akaike1974}
{Akaike}, H. 1974, IEEE Transactions on Automatic Control, 19, 716

\bibitem[{{Akerman} {et~al.}(2005){Akerman}, {Ellison}, {Pettini}, \&
  {Steidel}}]{Akerman2005}
{Akerman}, C.~J., {Ellison}, S.~L., {Pettini}, M., \& {Steidel}, C.~C. 2005,
  \aap, 440, 499, \dodoi{10.1051/0004-6361:20052947}

\bibitem[{{Asplund}(2005)}]{Asplund2005}
{Asplund}, M. 2005, \araa, 43, 481,
  \dodoi{10.1146/annurev.astro.42.053102.134001}

\bibitem[{{Asplund} {et~al.}(2009){Asplund}, {Grevesse}, {Sauval}, \&
  {Scott}}]{Asplund2009}
{Asplund}, M., {Grevesse}, N., {Sauval}, A.~J., \& {Scott}, P. 2009, \araa, 47,
  481, \dodoi{10.1146/annurev.astro.46.060407.145222}

\bibitem[{{Astropy Collaboration} {et~al.}(2013){Astropy Collaboration},
  {Robitaille}, {Tollerud}, {Greenfield}, {Droettboom}, {Bray}, {Aldcroft},
  {Davis}, {Ginsburg}, {Price-Whelan}, {Kerzendorf}, {Conley}, {Crighton},
  {Barbary}, {Muna}, {Ferguson}, {Grollier}, {Parikh}, {Nair}, {Unther},
  {Deil}, {Woillez}, {Conseil}, {Kramer}, {Turner}, {Singer}, {Fox}, {Weaver},
  {Zabalza}, {Edwards}, {Azalee Bostroem}, {Burke}, {Casey}, {Crawford},
  {Dencheva}, {Ely}, {Jenness}, {Labrie}, {Lim}, {Pierfederici}, {Pontzen},
  {Ptak}, {Refsdal}, {Servillat}, \& {Streicher}}]{ASTROPY}
{Astropy Collaboration}, {Robitaille}, T.~P., {Tollerud}, E.~J., {et~al.} 2013,
  \aap, 558, A33, \dodoi{10.1051/0004-6361/201322068}

\bibitem[{{Ba{\~n}ados} {et~al.}(2019){Ba{\~n}ados}, {Rauch}, {Decarli},
  {Farina}, {Hennawi}, {Mazzucchelli}, {Venemans}, {Walter}, {Simcoe},
  {Prochaska}, {Cooper}, {Davies}, \& {Chen}}]{Banados2019}
{Ba{\~n}ados}, E., {Rauch}, M., {Decarli}, R., {et~al.} 2019, \apj, 885, 59,
  \dodoi{10.3847/1538-4357/ab4129}

\bibitem[{{Barkana} \& {Loeb}(2001)}]{BarkanaLoeb2001}
{Barkana}, R., \& {Loeb}, A. 2001, \physrep, 349, 125,
  \dodoi{10.1016/S0370-1573(01)00019-9}

\bibitem[{{Bechtol} {et~al.}(2015){Bechtol}, {Drlica-Wagner}, {Balbinot},
  {Pieres}, {Simon}, {Yanny}, {Santiago}, {Wechsler}, {Frieman}, {Walker},
  {Williams}, {Rozo}, {Rykoff}, {Queiroz}, {Luque}, {Benoit-L{\'e}vy},
  {Tucker}, {Sevilla}, {Gruendl}, {da Costa}, {Fausti Neto}, {Maia}, {Abbott},
  {Allam}, {Armstrong}, {Bauer}, {Bernstein}, {Bernstein}, {Bertin}, {Brooks},
  {Buckley-Geer}, {Burke}, {Carnero Rosell}, {Castander}, {Covarrubias},
  {D'Andrea}, {DePoy}, {Desai}, {Diehl}, {Eifler}, {Estrada}, {Evrard},
  {Fernandez}, {Finley}, {Flaugher}, {Gaztanaga}, {Gerdes}, {Girardi},
  {Gladders}, {Gruen}, {Gutierrez}, {Hao}, {Honscheid}, {Jain}, {James},
  {Kent}, {Kron}, {Kuehn}, {Kuropatkin}, {Lahav}, {Li}, {Lin}, {Makler},
  {March}, {Marshall}, {Martini}, {Merritt}, {Miller}, {Miquel}, {Mohr},
  {Neilsen}, {Nichol}, {Nord}, {Ogando}, {Peoples}, {Petravick}, {Plazas},
  {Romer}, {Roodman}, {Sako}, {Sanchez}, {Scarpine}, {Schubnell}, {Smith},
  {Soares-Santos}, {Sobreira}, {Suchyta}, {Swanson}, {Tarle}, {Thaler},
  {Thomas}, {Wester}, {Zuntz}, \& {DES Collaboration}}]{Bechtol2015}
{Bechtol}, K., {Drlica-Wagner}, A., {Balbinot}, E., {et~al.} 2015, \apj, 807,
  50, \dodoi{10.1088/0004-637X/807/1/50}

\bibitem[{{Becker} {et~al.}(2011){Becker}, {Sargent}, {Rauch}, \&
  {Calverley}}]{Becker2011}
{Becker}, G.~D., {Sargent}, W.~L.~W., {Rauch}, M., \& {Calverley}, A.~P. 2011,
  \apj, 735, 93, \dodoi{10.1088/0004-637X/735/2/93}

\bibitem[{{Becker} {et~al.}(2019){Becker}, {Pettini}, {Rafelski}, {D'Odorico},
  {Boera}, {Christensen}, {Cupani}, {Ellison}, {Farina}, {Fumagalli},
  {L{\'o}pez}, {Neeleman}, {Ryan-Weber}, \& {Worseck}}]{Becker2019}
{Becker}, G.~D., {Pettini}, M., {Rafelski}, M., {et~al.} 2019, \apj, 883, 163,
  \dodoi{10.3847/1538-4357/ab3eb5}

\bibitem[{{Beers} \& {Carollo}(2008)}]{Beers2008}
{Beers}, T.~C., \& {Carollo}, D. 2008, in American Institute of Physics
  Conference Series, Vol. 990, First Stars III, ed. B.~W. {O'Shea} \&
  A.~{Heger}, 104--108, \dodoi{10.1063/1.2905513}

\bibitem[{{Beers} \& {Christlieb}(2005)}]{Beers2005}
{Beers}, T.~C., \& {Christlieb}, N. 2005, \araa, 43, 531,
  \dodoi{10.1146/annurev.astro.42.053102.134057}

\bibitem[{{Beers} {et~al.}(1985){Beers}, {Preston}, \& {Shectman}}]{Beers1985}
{Beers}, T.~C., {Preston}, G.~W., \& {Shectman}, S.~A. 1985, \aj, 90, 2089,
  \dodoi{10.1086/113917}

\bibitem[{{Beers} {et~al.}(1992){Beers}, {Preston}, \& {Shectman}}]{Beers1992}
---. 1992, \aj, 103, 1987, \dodoi{10.1086/116207}

\bibitem[{{Berg} {et~al.}(2022){Berg}, {Lehner}, {Howk}, {O'Meara}, {Schaye},
  {Straka}, {Cooksey}, {Tripp}, {Prochaska}, {Oppenheimer}, {Johnson},
  {Muzahid}, {Bordoloi}, {Werk}, {Fox}, {Katz}, {Wendt}, {Peeples}, {Ribaudo},
  \& {Tumlinson}}]{Berg2022}
{Berg}, M.~A., {Lehner}, N., {Howk}, J.~C., {et~al.} 2022, arXiv e-prints,
  arXiv:2204.13229.
\newblock \doarXiv{2204.13229}

\bibitem[{{Bernstein} {et~al.}(2015){Bernstein}, {Burles}, \&
  {Prochaska}}]{Bernstein2015}
{Bernstein}, R.~M., {Burles}, S.~M., \& {Prochaska}, J.~X. 2015, \pasp, 127,
  911, \dodoi{10.1086/683015}

\bibitem[{{Bird} {et~al.}(2013){Bird}, {Vogelsberger}, {Sijacki},
  {Zaldarriaga}, {Springel}, \& {Hernquist}}]{Bird2013}
{Bird}, S., {Vogelsberger}, M., {Sijacki}, D., {et~al.} 2013, \mnras, 429,
  3341, \dodoi{10.1093/mnras/sts590}

\bibitem[{{Bond}(1980)}]{Bond1980}
{Bond}, H.~E. 1980, The Astrophysical Journal Supplement Series, 44, 517,
  \dodoi{10.1086/190703}

\bibitem[{{Bosman} \& {Becker}(2015)}]{Bosman2015}
{Bosman}, S. E.~I., \& {Becker}, G.~D. 2015, \mnras, 452, 1105,
  \dodoi{10.1093/mnras/stv1336}

\bibitem[{{Bosman} {et~al.}(2022){Bosman}, {Davies}, {Becker}, {Keating},
  {Davies}, {Zhu}, {Eilers}, {D'Odorico}, {Bian}, {Bischetti}, {Cristiani},
  {Fan}, {Farina}, {Haehnelt}, {Hennawi}, {Kulkarni}, {Mesinger}, {Meyer},
  {Onoue}, {Pallottini}, {Qin}, {Ryan-Weber}, {Schindler}, {Walter}, {Wang}, \&
  {Yang}}]{Bosman2022}
{Bosman}, S. E.~I., {Davies}, F.~B., {Becker}, G.~D., {et~al.} 2022, \mnras,
  514, 55, \dodoi{10.1093/mnras/stac1046}

\bibitem[{{Bovill} \& {Ricotti}(2009)}]{Bovill2009}
{Bovill}, M.~S., \& {Ricotti}, M. 2009, \apj, 693, 1859,
  \dodoi{10.1088/0004-637X/693/2/1859}

\bibitem[{{Bovill} {et~al.}(2022){Bovill}, {Stiavelli}, {Wiggins}, {Ricotti},
  \& {Trenti}}]{Bovill2022}
{Bovill}, M.~S., {Stiavelli}, M., {Wiggins}, A.~I., {Ricotti}, M., \& {Trenti},
  M. 2022, arXiv e-prints, arXiv:2210.10190.
\newblock \doarXiv{2210.10190}

\bibitem[{{Bromm} {et~al.}(2002){Bromm}, {Coppi}, \&
  {Larson}}]{BrommCoppiLarson2002}
{Bromm}, V., {Coppi}, P.~S., \& {Larson}, R.~B. 2002, \apj, 564, 23,
  \dodoi{10.1086/323947}

\bibitem[{{Bromm} \& {Yoshida}(2011)}]{BrommYoshida2011}
{Bromm}, V., \& {Yoshida}, N. 2011, \araa, 49, 373,
  \dodoi{10.1146/annurev-astro-081710-102608}

\bibitem[{{Cayrel} {et~al.}(2004){Cayrel}, {Depagne}, {Spite}, {Hill}, {Spite},
  {Fran{\c c}ois}, {Plez}, {Beers}, {Primas}, {Andersen}, {Barbuy},
  {Bonifacio}, {Molaro}, \& {Nordstr{\"o}m}}]{Cayrel2004}
{Cayrel}, R., {Depagne}, E., {Spite}, M., {et~al.} 2004, \aap, 416, 1117,
  \dodoi{10.1051/0004-6361:20034074}

\bibitem[{{Chiti} {et~al.}(2018){Chiti}, {Frebel}, {Ji}, {Jerjen}, {Kim}, \&
  {Norris}}]{Chiti2018}
{Chiti}, A., {Frebel}, A., {Ji}, A.~P., {et~al.} 2018, \apj, 857, 74,
  \dodoi{10.3847/1538-4357/aab4fc}

\bibitem[{{Christlieb} {et~al.}(2008){Christlieb}, {Sch{\"o}rck}, {Frebel},
  {Beers}, {Wisotzki}, \& {Reimers}}]{Christlieb2008}
{Christlieb}, N., {Sch{\"o}rck}, T., {Frebel}, A., {et~al.} 2008, \aap, 484,
  721, \dodoi{10.1051/0004-6361:20078748}

\bibitem[{{Clark} {et~al.}(2011){Clark}, {Glover}, {Klessen}, \&
  {Bromm}}]{Clark2011}
{Clark}, P.~C., {Glover}, S.~C.~O., {Klessen}, R.~S., \& {Bromm}, V. 2011,
  \apj, 727, 110, \dodoi{10.1088/0004-637X/727/2/110}

\bibitem[{{Cooke} {et~al.}(2013){Cooke}, {Pettini}, {Jorgenson}, {Murphy},
  {Rudie}, \& {Steidel}}]{Cooke2013}
{Cooke}, R., {Pettini}, M., {Jorgenson}, R.~A., {et~al.} 2013, \mnras, 431,
  1625, \dodoi{10.1093/mnras/stt282}

\bibitem[{{Cooke} {et~al.}(2011{\natexlab{a}}){Cooke}, {Pettini}, {Steidel},
  {Rudie}, \& {Jorgenson}}]{Cooke2011a}
{Cooke}, R., {Pettini}, M., {Steidel}, C.~C., {Rudie}, G.~C., \& {Jorgenson},
  R.~A. 2011{\natexlab{a}}, \mnras, 412, 1047,
  \dodoi{10.1111/j.1365-2966.2010.17966.x}

\bibitem[{{Cooke} {et~al.}(2011{\natexlab{b}}){Cooke}, {Pettini}, {Steidel},
  {Rudie}, \& {Nissen}}]{Cooke2011b}
{Cooke}, R., {Pettini}, M., {Steidel}, C.~C., {Rudie}, G.~C., \& {Nissen},
  P.~E. 2011{\natexlab{b}}, \mnras, 417, 1534,
  \dodoi{10.1111/j.1365-2966.2011.19365.x}

\bibitem[{{Cooke} {et~al.}(2015){Cooke}, {Pettini}, \&
  {Jorgenson}}]{CookePettiniJorgenson2015}
{Cooke}, R.~J., {Pettini}, M., \& {Jorgenson}, R.~A. 2015, \apj, 800, 12,
  \dodoi{10.1088/0004-637X/800/1/12}

\bibitem[{{Cooke} {et~al.}(2016){Cooke}, {Pettini}, {Nollett}, \&
  {Jorgenson}}]{Cooke2016}
{Cooke}, R.~J., {Pettini}, M., {Nollett}, K.~M., \& {Jorgenson}, R. 2016, \apj,
  830, 148, \dodoi{10.3847/0004-637X/830/2/148}

\bibitem[{{Cooke} {et~al.}(2017){Cooke}, {Pettini}, \& {Steidel}}]{Cooke2017}
{Cooke}, R.~J., {Pettini}, M., \& {Steidel}, C.~C. 2017, \mnras, 467, 802,
  \dodoi{10.1093/mnras/stx037}

\bibitem[{{Dalton} {et~al.}(2012){Dalton}, {Trager}, {Abrams}, {Carter},
  {Bonifacio}, {Aguerri}, {MacIntosh}, {Evans}, {Lewis}, {Navarro}, {Agocs},
  {Dee}, {Rousset}, {Tosh}, {Middleton}, {Pragt}, {Terrett}, {Brock}, {Benn},
  {Verheijen}, {Cano Infantes}, {Bevil}, {Steele}, {Mottram}, {Bates},
  {Gribbin}, {Rey}, {Rodriguez}, {Delgado}, {Guinouard}, {Walton}, {Irwin},
  {Jagourel}, {Stuik}, {Gerlofsma}, {Roelfsma}, {Skillen}, {Ridings},
  {Balcells}, {Daban}, {Gouvret}, {Venema}, \& {Girard}}]{Dalton2012}
{Dalton}, G., {Trager}, S.~C., {Abrams}, D.~C., {et~al.} 2012, in Society of
  Photo-Optical Instrumentation Engineers (SPIE) Conference Series, Vol. 8446,
  Ground-based and Airborne Instrumentation for Astronomy IV, ed. I.~S.
  {McLean}, S.~K. {Ramsay}, \& H.~{Takami}, 84460P, \dodoi{10.1117/12.925950}

\bibitem[{{de Jong} {et~al.}(2012){de Jong}, {Bellido-Tirado}, {Chiappini},
  {Depagne}, {Haynes}, {Johl}, {Schnurr}, {Schwope}, {Walcher}, {Dionies},
  {Haynes}, {Kelz}, {Kitaura}, {Lamer}, {Minchev}, {M{\"u}ller}, {Nuza},
  {Olaya}, {Piffl}, {Popow}, {Steinmetz}, {Ural}, {Williams}, {Winkler},
  {Wisotzki}, {Ansorge}, {Banerji}, {Gonzalez Solares}, {Irwin}, {Kennicutt},
  {King}, {McMahon}, {Koposov}, {Parry}, {Sun}, {Walton}, {Finger}, {Iwert},
  {Krumpe}, {Lizon}, {Vincenzo}, {Amans}, {Bonifacio}, {Cohen}, {Francois},
  {Jagourel}, {Mignot}, {Royer}, {Sartoretti}, {Bender}, {Grupp}, {Hess},
  {Lang-Bardl}, {Muschielok}, {B{\"o}hringer}, {Boller}, {Bongiorno}, {Brusa},
  {Dwelly}, {Merloni}, {Nandra}, {Salvato}, {Pragt}, {Navarro}, {Gerlofsma},
  {Roelfsema}, {Dalton}, {Middleton}, {Tosh}, {Boeche}, {Caffau}, {Christlieb},
  {Grebel}, {Hansen}, {Koch}, {Ludwig}, {Quirrenbach}, {Sbordone}, {Seifert},
  {Thimm}, {Trifonov}, {Helmi}, {Trager}, {Feltzing}, {Korn}, \&
  {Boland}}]{deJong2012}
{de Jong}, R.~S., {Bellido-Tirado}, O., {Chiappini}, C., {et~al.} 2012, in
  Society of Photo-Optical Instrumentation Engineers (SPIE) Conference Series,
  Vol. 8446, Ground-based and Airborne Instrumentation for Astronomy IV, ed.
  I.~S. {McLean}, S.~K. {Ramsay}, \& H.~{Takami}, 84460T,
  \dodoi{10.1117/12.926239}

\bibitem[{{DESI Collaboration} {et~al.}(2016){DESI Collaboration}, {Aghamousa},
  {Aguilar}, {Ahlen}, {Alam}, {Allen}, {Allende Prieto}, {Annis}, {Bailey},
  {Balland}, {Ballester}, {Baltay}, {Beaufore}, {Bebek}, {Beers}, {Bell},
  {Bernal}, {Besuner}, {Beutler}, {Blake}, {Bleuler}, {Blomqvist}, {Blum},
  {Bolton}, {Briceno}, {Brooks}, {Brownstein}, {Buckley-Geer}, {Burden},
  {Burtin}, {Busca}, {Cahn}, {Cai}, {Cardiel-Sas}, {Carlberg}, {Carton},
  {Casas}, {Castander}, {Cervantes-Cota}, {Claybaugh}, {Close}, {Coker},
  {Cole}, {Comparat}, {Cooper}, {Cousinou}, {Crocce}, {Cuby}, {Cunningham},
  {Davis}, {Dawson}, {de la Macorra}, {De Vicente}, {Delubac}, {Derwent},
  {Dey}, {Dhungana}, {Ding}, {Doel}, {Duan}, {Ealet}, {Edelstein},
  {Eftekharzadeh}, {Eisenstein}, {Elliott}, {Escoffier}, {Evatt}, {Fagrelius},
  {Fan}, {Fanning}, {Farahi}, {Farihi}, {Favole}, {Feng}, {Fernandez},
  {Findlay}, {Finkbeiner}, {Fitzpatrick}, {Flaugher}, {Flender}, {Font-Ribera},
  {Forero-Romero}, {Fosalba}, {Frenk}, {Fumagalli}, {Gaensicke}, {Gallo},
  {Garcia-Bellido}, {Gaztanaga}, {Pietro Gentile Fusillo}, {Gerard},
  {Gershkovich}, {Giannantonio}, {Gillet}, {Gonzalez-de-Rivera},
  {Gonzalez-Perez}, {Gott}, {Graur}, {Gutierrez}, {Guy}, {Habib}, {Heetderks},
  {Heetderks}, {Heitmann}, {Hellwing}, {Herrera}, {Ho}, {Holland}, {Honscheid},
  {Huff}, {Hutchinson}, {Huterer}, {Hwang}, {Illa Laguna}, {Ishikawa},
  {Jacobs}, {Jeffrey}, {Jelinsky}, {Jennings}, {Jiang}, {Jimenez}, {Johnson},
  {Joyce}, {Jullo}, {Juneau}, {Kama}, {Karcher}, {Karkar}, {Kehoe}, {Kennamer},
  {Kent}, {Kilbinger}, {Kim}, {Kirkby}, {Kisner}, {Kitanidis}, {Kneib},
  {Koposov}, {Kovacs}, {Koyama}, {Kremin}, {Kron}, {Kronig}, {Kueter-Young},
  {Lacey}, {Lafever}, {Lahav}, {Lambert}, {Lampton}, {Landriau}, {Lang},
  {Lauer}, {Le Goff}, {Le Guillou}, {Le Van Suu}, {Lee}, {Lee}, {Leitner},
  {Lesser}, {Levi}, {L'Huillier}, {Li}, {Liang}, {Lin}, {Linder}, {Loebman},
  {Luki{\'c}}, {Ma}, {MacCrann}, {Magneville}, {Makarem}, {Manera}, {Manser},
  {Marshall}, {Martini}, {Massey}, {Matheson}, {McCauley}, {McDonald},
  {McGreer}, {Meisner}, {Metcalfe}, {Miller}, {Miquel}, {Moustakas}, {Myers},
  {Naik}, {Newman}, {Nichol}, {Nicola}, {Nicolati da Costa}, {Nie}, {Niz},
  {Norberg}, {Nord}, {Norman}, {Nugent}, {O'Brien}, {Oh}, {Olsen}, {Padilla},
  {Padmanabhan}, {Padmanabhan}, {Palanque-Delabrouille}, {Palmese},
  {Pappalardo}, {P{\^a}ris}, {Park}, {Patej}, {Peacock}, {Peiris}, {Peng},
  {Percival}, {Perruchot}, {Pieri}, {Pogge}, {Pollack}, {Poppett}, {Prada},
  {Prakash}, {Probst}, {Rabinowitz}, {Raichoor}, {Ree}, {Refregier}, {Regal},
  {Reid}, {Reil}, {Rezaie}, {Rockosi}, {Roe}, {Ronayette}, {Roodman}, {Ross},
  {Ross}, {Rossi}, {Rozo}, {Ruhlmann-Kleider}, {Rykoff}, {Sabiu}, {Samushia},
  {Sanchez}, {Sanchez}, {Schlegel}, {Schneider}, {Schubnell}, {Secroun},
  {Seljak}, {Seo}, {Serrano}, {Shafieloo}, {Shan}, {Sharples}, {Sholl},
  {Shourt}, {Silber}, {Silva}, {Sirk}, {Slosar}, {Smith}, {Smoot}, {Som},
  {Song}, {Sprayberry}, {Staten}, {Stefanik}, {Tarle}, {Sien Tie}, {Tinker},
  {Tojeiro}, {Valdes}, {Valenzuela}, {Valluri}, {Vargas-Magana}, {Verde},
  {Walker}, {Wang}, {Wang}, {Weaver}, {Weaverdyck}, {Wechsler}, {Weinberg},
  {White}, {Yang}, {Yeche}, {Zhang}, {Zhao}, {Zheng}, {Zhou}, {Zhou}, {Zhu},
  {Zou}, \& {Zu}}]{DESI2016}
{DESI Collaboration}, {Aghamousa}, A., {Aguilar}, J., {et~al.} 2016, arXiv
  e-prints, arXiv:1611.00036.
\newblock \doarXiv{1611.00036}

\bibitem[{{D'Odorico} {et~al.}(2022){D'Odorico}, {Finlator}, {Cristiani},
  {Cupani}, {Perrotta}, {Calura}, {C{\`e}nturion}, {Becker}, {Berg}, {Lopez},
  {Ellison}, \& {Pomante}}]{Dodorico2022}
{D'Odorico}, V., {Finlator}, K., {Cristiani}, S., {et~al.} 2022, \mnras, 512,
  2389, \dodoi{10.1093/mnras/stac545}

\bibitem[{{D'Odorico} {et~al.}(2023){D'Odorico}, {Banados}, {Becker},
  {Bischetti}, {Bosman}, {Cupani}, {Davies}, {Farina}, {Ferrara}, {Feruglio},
  {Mazzucchelli}, {Ryan-Weber}, {Schindler}, {Sodini}, {Venemans}, {Walter},
  {Chen}, {Lai}, {Zhu}, {Bian}, {Campo}, {Carniani}, {Cristiani}, {Davies},
  {Decarli}, {Drake}, {Eilers}, {Fan}, {Gaikwad}, {Gallerani}, {Greig},
  {Haehnelt}, {Hennawi}, {Keating}, {Kulkarni}, {Mesinger}, {Meyer},
  {Neeleman}, {Onoue}, {Pallottini}, {Qin}, {Rojas-Ruiz}, {Satyavolu},
  {Sebastian}, {Tripodi}, {Wang}, {Wolfson}, {Yang}, \&
  {Zanchettin}}]{Dodorico2023}
{D'Odorico}, V., {Banados}, E., {Becker}, G.~D., {et~al.} 2023, arXiv e-prints,
  arXiv:2305.05053, \dodoi{10.48550/arXiv.2305.05053}

\bibitem[{{Dutta} {et~al.}(2014){Dutta}, {Srianand}, {Rahmani}, {Petitjean},
  {Noterdaeme}, \& {Ledoux}}]{Dutta2014}
{Dutta}, R., {Srianand}, R., {Rahmani}, H., {et~al.} 2014, \mnras, 440, 307,
  \dodoi{10.1093/mnras/stu260}

\bibitem[{{Ellison} {et~al.}(2010){Ellison}, {Prochaska}, {Hennawi}, {Lopez},
  {Usher}, {Wolfe}, {Russell}, \& {Benn}}]{Ellison2010}
{Ellison}, S.~L., {Prochaska}, J.~X., {Hennawi}, J., {et~al.} 2010, \mnras,
  406, 1435, \dodoi{10.1111/j.1365-2966.2010.16780.x}

\bibitem[{{Erni} {et~al.}(2006){Erni}, {Richter}, {Ledoux}, \&
  {Petitjean}}]{Erni2006}
{Erni}, P., {Richter}, P., {Ledoux}, C., \& {Petitjean}, P. 2006, \aap, 451,
  19, \dodoi{10.1051/0004-6361:20054328}

\bibitem[{{Foreman-Mackey}(2016)}]{CORNER}
{Foreman-Mackey}, D. 2016, The Journal of Open Source Software, 1, 24,
  \dodoi{10.21105/joss.00024}

\bibitem[{{Foreman-Mackey} {et~al.}(2013){Foreman-Mackey}, {Hogg}, {Lang}, \&
  {Goodman}}]{EMCEE}
{Foreman-Mackey}, D., {Hogg}, D.~W., {Lang}, D., \& {Goodman}, J. 2013, \pasp,
  125, 306, \dodoi{10.1086/670067}

\bibitem[{{Fumagalli} {et~al.}(2011){Fumagalli}, {O'Meara}, \&
  {Prochaska}}]{Fumagalli2012}
{Fumagalli}, M., {O'Meara}, J.~M., \& {Prochaska}, J.~X. 2011, Science, 334,
  1245, \dodoi{10.1126/science.1213581}

\bibitem[{{Fumagalli} {et~al.}(2015){Fumagalli}, {O'Meara}, {Prochaska},
  {Rafelski}, \& {Kanekar}}]{Fumagalli2015}
{Fumagalli}, M., {O'Meara}, J.~M., {Prochaska}, J.~X., {Rafelski}, M., \&
  {Kanekar}, N. 2015, MNRAS, 446, 3178, \dodoi{10.1093/mnras/stu2325}

\bibitem[{{Fumagalli} {et~al.}(2017){Fumagalli}, {Mackenzie}, {Trayford},
  {Theuns}, {Cantalupo}, {Christensen}, {Fynbo}, {M{\o}ller}, {O'Meara},
  {Prochaska}, {Rafelski}, \& {Shanks}}]{Fumagalli2017}
{Fumagalli}, M., {Mackenzie}, R., {Trayford}, J., {et~al.} 2017, MNRAS, 471,
  3686, \dodoi{10.1093/mnras/stx1896}

\bibitem[{{Gardner} {et~al.}(2009){Gardner}, {Mather}, {Clampin}, {Doyon},
  {Flanagan}, {Franx}, {Greenhouse}, {Hammel}, {Hutchings}, {Jakobsen},
  {Lilly}, {Lunine}, {McCaughrean}, {Mountain}, {Rieke}, {Rieke}, {Sonneborn},
  {Stiavelli}, {Windhorst}, \& {Wright}}]{Gardner2009}
{Gardner}, J.~P., {Mather}, J.~C., {Clampin}, M., {et~al.} 2009, in
  Astrophysics and Space Science Proceedings, Vol.~10, Astrophysics in the Next
  Decade, 1, \dodoi{10.1007/978-1-4020-9457-6_1}

\bibitem[{{Gilmozzi} \& {Spyromilio}(2007)}]{eltpaper}
{Gilmozzi}, R., \& {Spyromilio}, J. 2007, The Messenger, 127, 11

\bibitem[{{Greif} {et~al.}(2010){Greif}, {Glover}, {Bromm}, \&
  {Klessen}}]{Greif2010}
{Greif}, T.~H., {Glover}, S.~C.~O., {Bromm}, V., \& {Klessen}, R.~S. 2010,
  \apj, 716, 510, \dodoi{10.1088/0004-637X/716/1/510}

\bibitem[{{Haehnelt} {et~al.}(1998){Haehnelt}, {Steinmetz}, \&
  {Rauch}}]{Haehnelt1998}
{Haehnelt}, M.~G., {Steinmetz}, M., \& {Rauch}, M. 1998, \apj, 495, 647,
  \dodoi{10.1086/305323}

\bibitem[{{Heger} \& {Woosley}(2010)}]{HegerWoosley2010}
{Heger}, A., \& {Woosley}, S.~E. 2010, \apj, 724, 341,
  \dodoi{10.1088/0004-637X/724/1/341}

\bibitem[{{Hirano} {et~al.}(2014){Hirano}, {Hosokawa}, {Yoshida}, {Umeda},
  {Omukai}, {Chiaki}, \& {Yorke}}]{Hirano2014}
{Hirano}, S., {Hosokawa}, T., {Yoshida}, N., {et~al.} 2014, \apj, 781, 60,
  \dodoi{10.1088/0004-637X/781/2/60}

\bibitem[{{Howes} {et~al.}(2016){Howes}, {Asplund}, {Keller}, {Casey}, {Yong},
  {Lind}, {Frebel}, {Hays}, {Alves-Brito}, {Bessell}, {Casagrande}, {Marino},
  {Nataf}, {Owen}, {Da Costa}, {Schmidt}, \& {Tisserand}}]{Howes2016}
{Howes}, L.~M., {Asplund}, M., {Keller}, S.~C., {et~al.} 2016, \mnras, 460,
  884, \dodoi{10.1093/mnras/stw1004}

\bibitem[{{Hunter}(2007)}]{MATPLOTLIB}
{Hunter}, J.~D. 2007, Computing in Science and Engineering, 9, 90,
  \dodoi{10.1109/MCSE.2007.55}

\bibitem[{{Johns} {et~al.}(2012){Johns}, {McCarthy}, {Raybould}, {Bouchez},
  {Farahani}, {Filgueira}, {Jacoby}, {Shectman}, \& {Sheehan}}]{gmtpaper}
{Johns}, M., {McCarthy}, P., {Raybould}, K., {et~al.} 2012, in Society of
  Photo-Optical Instrumentation Engineers (SPIE) Conference Series, Vol. 8444,
  Ground-based and Airborne Telescopes IV, ed. L.~M. {Stepp}, R.~{Gilmozzi}, \&
  H.~J. {Hall}, 84441H, \dodoi{10.1117/12.926716}

\bibitem[{{Klitsch} {et~al.}(2021){Klitsch}, {P{\'e}roux}, {Zwaan}, {De Cia},
  {Ledoux}, \& {Lopez}}]{Klitsch2021}
{Klitsch}, A., {P{\'e}roux}, C., {Zwaan}, M.~A., {et~al.} 2021, MNRAS, 506,
  514, \dodoi{10.1093/mnras/stab1668}

\bibitem[{{Krogager} {et~al.}(2017){Krogager}, {M{\o}ller}, {Fynbo}, \&
  {Noterdaeme}}]{Krogager2017}
{Krogager}, J.~K., {M{\o}ller}, P., {Fynbo}, J.~P.~U., \& {Noterdaeme}, P.
  2017, MNRAS, 469, 2959, \dodoi{10.1093/mnras/stx1011}

\bibitem[{{Limongi} \& {Chieffi}(2018)}]{ChieffiLimongi2018}
{Limongi}, M., \& {Chieffi}, A. 2018, ArXiv e-prints.
\newblock \doarXiv{1805.09640}

\bibitem[{{Lodders}(2019)}]{Lodders2019}
{Lodders}, K. 2019, arXiv e-prints, arXiv:1912.00844,
  \dodoi{10.48550/arXiv.1912.00844}

\bibitem[{{Lofthouse} {et~al.}(2022){Lofthouse}, {Fumagalli}, {Fossati},
  {Dutta}, {Galbiati}, {Battaia}, {Cantalupo}, {Christensen}, {Cooke},
  {Longobardi}, {Murphy}, \& {Prochaska}}]{Lofthouse2022}
{Lofthouse}, E.~K., {Fumagalli}, M., {Fossati}, M., {et~al.} 2022, \mnras,
  \dodoi{10.1093/mnras/stac3089}

\bibitem[{{Mackenzie} {et~al.}(2019){Mackenzie}, {Fumagalli}, {Theuns},
  {Hatton}, {Garel}, {Cantalupo}, {Christensen}, {Fynbo}, {Kanekar},
  {M{\o}ller}, {O'Meara}, {Prochaska}, {Rafelski}, {Shanks}, \&
  {Trayford}}]{Mackenzie2019}
{Mackenzie}, R., {Fumagalli}, M., {Theuns}, T., {et~al.} 2019, MNRAS, 487,
  5070, \dodoi{10.1093/mnras/stz1501}

\bibitem[{{McWilliam} {et~al.}(1995){McWilliam}, {Preston}, {Sneden}, \&
  {Shectman}}]{McWilliam1995}
{McWilliam}, A., {Preston}, G.~W., {Sneden}, C., \& {Shectman}, S. 1995, \aj,
  109, 2736, \dodoi{10.1086/117485}

\bibitem[{{M{\o}ller} {et~al.}(2018){M{\o}ller}, {Christensen}, {Zwaan},
  {Kanekar}, {Prochaska}, {Rhodin}, {Dessauges-Zavadsky}, {Fynbo}, {Neeleman},
  \& {Zafar}}]{Moller2018}
{M{\o}ller}, P., {Christensen}, L., {Zwaan}, M.~A., {et~al.} 2018, MNRAS, 474,
  4039, \dodoi{10.1093/mnras/stx2845}

\bibitem[{{Mu{\~n}oz} {et~al.}(2009){Mu{\~n}oz}, {Madau}, {Loeb}, \&
  {Diemand}}]{Munoz2009}
{Mu{\~n}oz}, J.~A., {Madau}, P., {Loeb}, A., \& {Diemand}, J. 2009, \mnras,
  400, 1593, \dodoi{10.1111/j.1365-2966.2009.15562.x}

\bibitem[{{Neeleman} {et~al.}(2018){Neeleman}, {Kanekar}, {Prochaska},
  {Christensen}, {Dessauges-Zavadsky}, {Fynbo}, {M{\o}ller}, \&
  {Zwaan}}]{Neeleman2018}
{Neeleman}, M., {Kanekar}, N., {Prochaska}, J.~X., {et~al.} 2018, ApJl, 856,
  L12, \dodoi{10.3847/2041-8213/aab5b1}

\bibitem[{{Norris} {et~al.}(2010){Norris}, {Yong}, {Gilmore}, \&
  {Wyse}}]{Norris2010}
{Norris}, J.~E., {Yong}, D., {Gilmore}, G., \& {Wyse}, R. F.~G. 2010, \apj,
  711, 350, \dodoi{10.1088/0004-637X/711/1/350}

\bibitem[{{Noterdaeme} {et~al.}(2021){Noterdaeme}, {Balashev}, {Ledoux},
  {Duchoquet}, {L{\'o}pez}, {Telikova}, {Boiss{\'e}}, {Krogager}, {De Cia}, \&
  {Bergeron}}]{Noterdaeme2021}
{Noterdaeme}, P., {Balashev}, S., {Ledoux}, C., {et~al.} 2021, arXiv e-prints,
  arXiv:2105.00697.
\newblock \doarXiv{2105.00697}

\bibitem[{{Nu{\~n}ez} {et~al.}(2022){Nu{\~n}ez}, {Kirby}, \&
  {Steidel}}]{HazeNunez2022}
{Nu{\~n}ez}, E.~H., {Kirby}, E.~N., \& {Steidel}, C.~C. 2022, \apj, 927, 64,
  \dodoi{10.3847/1538-4357/ac470e}

\bibitem[{{P{\'e}roux} {et~al.}(2011){P{\'e}roux}, {Bouch{\'e}}, {Kulkarni},
  {York}, \& {Vladilo}}]{Peroux2011}
{P{\'e}roux}, C., {Bouch{\'e}}, N., {Kulkarni}, V.~P., {York}, D.~G., \&
  {Vladilo}, G. 2011, MNRAS, 410, 2251,
  \dodoi{10.1111/j.1365-2966.2010.17597.x}

\bibitem[{{P{\'e}roux} {et~al.}(2012){P{\'e}roux}, {Bouch{\'e}}, {Kulkarni},
  {York}, \& {Vladilo}}]{Peroux2012}
---. 2012, MNRAS, 419, 3060, \dodoi{10.1111/j.1365-2966.2011.19947.x}

\bibitem[{{P{\'e}roux} \& {Howk}(2020)}]{Peroux2020}
{P{\'e}roux}, C., \& {Howk}, J.~C. 2020, \araa, 58, 363,
  \dodoi{10.1146/annurev-astro-021820-120014}

\bibitem[{{Pettini} {et~al.}(1990){Pettini}, {Boksenberg}, \&
  {Hunstead}}]{Pettini1990}
{Pettini}, M., {Boksenberg}, A., \& {Hunstead}, R.~W. 1990, \apj, 348, 48,
  \dodoi{10.1086/168212}

\bibitem[{{Pettini} {et~al.}(1997){Pettini}, {King}, {Smith}, \&
  {Hunstead}}]{Pettini1997}
{Pettini}, M., {King}, D.~L., {Smith}, L.~J., \& {Hunstead}, R.~W. 1997, \apj,
  478, 536, \dodoi{10.1086/303826}

\bibitem[{{Pettini} {et~al.}(2008){Pettini}, {Zych}, {Steidel}, \&
  {Chaffee}}]{Pettini2008}
{Pettini}, M., {Zych}, B.~J., {Steidel}, C.~C., \& {Chaffee}, F.~H. 2008,
  \mnras, 385, 2011, \dodoi{10.1111/j.1365-2966.2008.12951.x}

\bibitem[{{Pieri} {et~al.}(2016){Pieri}, {Bonoli}, {Chaves-Montero},
  {P{\^a}ris}, {Fumagalli}, {Bolton}, {Viel}, {Noterdaeme},
  {Miralda-Escud{\'e}}, {Busca}, {Rahmani}, {Peroux}, {Font-Ribera}, \&
  {Trager}}]{Pieri2016}
{Pieri}, M.~M., {Bonoli}, S., {Chaves-Montero}, J., {et~al.} 2016, in
  SF2A-2016: Proceedings of the Annual meeting of the French Society of
  Astronomy and Astrophysics, ed. C.~{Reyl{\'e}}, J.~{Richard},
  L.~{Cambr{\'e}sy}, M.~{Deleuil}, E.~{P{\'e}contal}, L.~{Tresse}, \&
  I.~{Vauglin}, 259--266.
\newblock \doarXiv{1611.09388}

\bibitem[{{Planck Collaboration} {et~al.}(2018){Planck Collaboration},
  {Aghanim}, {Akrami}, {Ashdown}, {Aumont}, {Baccigalupi}, {Ballardini},
  {Banday}, {Barreiro}, {Bartolo}, {Basak}, {Battye}, {Benabed}, {Bernard},
  {Bersanelli}, {Bielewicz}, {Bock}, {Bond}, {Borrill}, {Bouchet}, {Boulanger},
  {Bucher}, {Burigana}, {Butler}, {Calabrese}, {Cardoso}, {Carron},
  {Challinor}, {Chiang}, {Chluba}, {Colombo}, {Combet}, {Contreras}, {Crill},
  {Cuttaia}, {de Bernardis}, {de Zotti}, {Delabrouille}, {Delouis}, {Di
  Valentino}, {Diego}, {Dor{\'e}}, {Douspis}, {Ducout}, {Dupac}, {Dusini},
  {Efstathiou}, {Elsner}, {En{\ss}lin}, {Eriksen}, {Fantaye}, {Farhang},
  {Fergusson}, {Fernandez-Cobos}, {Finelli}, {Forastieri}, {Frailis},
  {Fraisse}, {Franceschi}, {Frolov}, {Galeotta}, {Galli}, {Ganga},
  {G{\'e}nova-Santos}, {Gerbino}, {Ghosh}, {Gonz{\'a}lez-Nuevo}, {G{\'o}rski},
  {Gratton}, {Gruppuso}, {Gudmundsson}, {Hamann}, {Handley}, {Hansen},
  {Herranz}, {Hildebrandt}, {Hivon}, {Huang}, {Jaffe}, {Jones}, {Karakci},
  {Keih{\"a}nen}, {Keskitalo}, {Kiiveri}, {Kim}, {Kisner}, {Knox},
  {Krachmalnicoff}, {Kunz}, {Kurki-Suonio}, {Lagache}, {Lamarre}, {Lasenby},
  {Lattanzi}, {Lawrence}, {Le Jeune}, {Lemos}, {Lesgourgues}, {Levrier},
  {Lewis}, {Liguori}, {Lilje}, {Lilley}, {Lindholm}, {L{\'o}pez-Caniego},
  {Lubin}, {Ma}, {Mac{\'\i}as-P{\'e}rez}, {Maggio}, {Maino}, {Mandolesi},
  {Mangilli}, {Marcos-Caballero}, {Maris}, {Martin}, {Martinelli},
  {Mart{\'\i}nez-Gonz{\'a}lez}, {Matarrese}, {Mauri}, {McEwen}, {Meinhold},
  {Melchiorri}, {Mennella}, {Migliaccio}, {Millea}, {Mitra},
  {Miville-Desch{\^e}nes}, {Molinari}, {Montier}, {Morgante}, {Moss}, {Natoli},
  {N{\o}rgaard-Nielsen}, {Pagano}, {Paoletti}, {Partridge}, {Patanchon},
  {Peiris}, {Perrotta}, {Pettorino}, {Piacentini}, {Polastri}, {Polenta},
  {Puget}, {Rachen}, {Reinecke}, {Remazeilles}, {Renzi}, {Rocha}, {Rosset},
  {Roudier}, {Rubi{\~n}o-Mart{\'\i}n}, {Ruiz-Granados}, {Salvati}, {Sandri},
  {Savelainen}, {Scott}, {Shellard}, {Sirignano}, {Sirri}, {Spencer},
  {Sunyaev}, {Suur-Uski}, {Tauber}, {Tavagnacco}, {Tenti}, {Toffolatti},
  {Tomasi}, {Trombetti}, {Valenziano}, {Valiviita}, {Van Tent}, {Vibert},
  {Vielva}, {Villa}, {Vittorio}, {Wand elt}, {Wehus}, {White}, {White},
  {Zacchei}, \& {Zonca}}]{Planck2018}
{Planck Collaboration}, {Aghanim}, N., {Akrami}, Y., {et~al.} 2018, arXiv
  e-prints, arXiv:1807.06209.
\newblock \doarXiv{1807.06209}

\bibitem[{{Pontzen} {et~al.}(2008){Pontzen}, {Governato}, {Pettini}, {Booth},
  {Stinson}, {Wadsley}, {Brooks}, {Quinn}, \& {Haehnelt}}]{Pontzen2008}
{Pontzen}, A., {Governato}, F., {Pettini}, M., {et~al.} 2008, \mnras, 390,
  1349, \dodoi{10.1111/j.1365-2966.2008.13782.x}

\bibitem[{{Rafelski} {et~al.}(2014){Rafelski}, {Neeleman}, {Fumagalli},
  {Wolfe}, \& {Prochaska}}]{Rafelski2014}
{Rafelski}, M., {Neeleman}, M., {Fumagalli}, M., {Wolfe}, A.~M., \&
  {Prochaska}, J.~X. 2014, \apj, 782, L29, \dodoi{10.1088/2041-8205/782/2/L29}

\bibitem[{{Rafelski} {et~al.}(2012){Rafelski}, {Wolfe}, {Prochaska},
  {Neeleman}, \& {Mendez}}]{Rafelski2012}
{Rafelski}, M., {Wolfe}, A.~M., {Prochaska}, J.~X., {Neeleman}, M., \&
  {Mendez}, A.~J. 2012, \apj, 755, 89, \dodoi{10.1088/0004-637X/755/2/89}

\bibitem[{{Rahmati} \& {Schaye}(2014)}]{Rahmati2014}
{Rahmati}, A., \& {Schaye}, J. 2014, \mnras, 438, 529,
  \dodoi{10.1093/mnras/stt2235}

\bibitem[{{Robert} {et~al.}(2019){Robert}, {Murphy}, {O'Meara}, {Crighton}, \&
  {Fumagalli}}]{Robert2019}
{Robert}, P.~F., {Murphy}, M.~T., {O'Meara}, J.~M., {Crighton}, N. H.~M., \&
  {Fumagalli}, M. 2019, \mnras, 483, 2736, \dodoi{10.1093/mnras/sty3287}

\bibitem[{{Roederer} {et~al.}(2014){Roederer}, {Preston}, {Thompson},
  {Shectman}, {Sneden}, {Burley}, \& {Kelson}}]{Roederer2014}
{Roederer}, I.~U., {Preston}, G.~W., {Thompson}, I.~B., {et~al.} 2014, \aj,
  147, 136, \dodoi{10.1088/0004-6256/147/6/136}

\bibitem[{{Rossi} {et~al.}(2023){Rossi}, {Salvadori}, {Sk{\'u}lad{\'o}ttir}, \&
  {Vanni}}]{Rossi2023}
{Rossi}, M., {Salvadori}, S., {Sk{\'u}lad{\'o}ttir}, {\'A}., \& {Vanni}, I.
  2023, arXiv e-prints, arXiv:2302.10210.
\newblock \doarXiv{2302.10210}

\bibitem[{{Ryan} {et~al.}(1996){Ryan}, {Norris}, \& {Beers}}]{Ryan1996}
{Ryan}, S.~G., {Norris}, J.~E., \& {Beers}, T.~C. 1996, \apj, 471, 254,
  \dodoi{10.1086/177967}

\bibitem[{{Ryan} {et~al.}(1991){Ryan}, {Norris}, \& {Bessell}}]{Ryan1991}
{Ryan}, S.~G., {Norris}, J.~E., \& {Bessell}, M.~S. 1991, \aj, 102, 303,
  \dodoi{10.1086/115878}

\bibitem[{{Saccardi} {et~al.}(2023){Saccardi}, {Salvadori}, {D'Odorico},
  {Cupani}, {Fumagalli}, {Berg}, {Becker}, {Ellison}, \&
  {Lopez}}]{Saccardi2023}
{Saccardi}, A., {Salvadori}, S., {D'Odorico}, V., {et~al.} 2023, \apj, 948, 35,
  \dodoi{10.3847/1538-4357/acc39f}

\bibitem[{{Salvadori} \& {Ferrara}(2009)}]{Salvadori2009}
{Salvadori}, S., \& {Ferrara}, A. 2009, \mnras, 395, L6,
  \dodoi{10.1111/j.1745-3933.2009.00627.x}

\bibitem[{{Schaerer}(2003)}]{Schaerer2003}
{Schaerer}, D. 2003, \aap, 397, 527, \dodoi{10.1051/0004-6361:20021525}

\bibitem[{{Senchyna} {et~al.}(2019){Senchyna}, {Stark}, {Chevallard},
  {Charlot}, {Jones}, \& {Vidal-Garc{\'\i}a}}]{Senchyna2019}
{Senchyna}, P., {Stark}, D.~P., {Chevallard}, J., {et~al.} 2019, \mnras, 488,
  3492, \dodoi{10.1093/mnras/stz1907}

\bibitem[{{Simcoe} {et~al.}(2012){Simcoe}, {Sullivan}, {Cooksey}, {Kao},
  {Matejek}, \& {Burgasser}}]{Simcoe2012}
{Simcoe}, R.~A., {Sullivan}, P.~W., {Cooksey}, K.~L., {et~al.} 2012, \nat, 492,
  79, \dodoi{10.1038/nature11612}

\bibitem[{{Simon}(2019)}]{Simon2019}
{Simon}, J.~D. 2019, \araa, 57, 375,
  \dodoi{10.1146/annurev-astro-091918-104453}

\bibitem[{{Skidmore} {et~al.}(2015){Skidmore}, {TMT International Science
  Development Teams}, \& {Science Advisory Committee}}]{tmtpaper}
{Skidmore}, W., {TMT International Science Development Teams}, \& {Science
  Advisory Committee}, T. 2015, Research in Astronomy and Astrophysics, 15,
  1945, \dodoi{10.1088/1674-4527/15/12/001}

\bibitem[{{Sk{\'u}lad{\'o}ttir} {et~al.}(2023){Sk{\'u}lad{\'o}ttir}, {Puls},
  {Amarsi}, {Battaglia}, {Buder}, {Campbell}, {Cardona-Barrero}, {Christlieb},
  {Feuillet}, {Gelli}, {Hansen}, {Hill}, {Ibata}, {Jablonka}, {Kacharov},
  {Karakas}, {Koch-Hansen}, {Lind}, {Lombardo}, {Lucchesi}, {Lugaro}, {Martin},
  {Massari}, {Nordlander}, {Reichert}, {Rossi}, {Ruiter}, {Salvadori},
  {Seitenzahl}, {Tolstoy}, {Xylakis-Dornbusch}, \& {Youakim}}]{Skuladottir2023}
{Sk{\'u}lad{\'o}ttir}, {\'A}., {Puls}, A.~A., {Amarsi}, A.~M., {et~al.} 2023,
  The Messenger, 190, 19, \dodoi{10.18727/0722-6691/5304}

\bibitem[{{Stacy} {et~al.}(2016){Stacy}, {Bromm}, \& {Lee}}]{Stacy2016}
{Stacy}, A., {Bromm}, V., \& {Lee}, A.~T. 2016, \mnras, 462, 1307,
  \dodoi{10.1093/mnras/stw1728}

\bibitem[{{Starkenburg} {et~al.}(2017){Starkenburg}, {Martin}, {Youakim},
  {Aguado}, {Allende Prieto}, {Arentsen}, {Bernard}, {Bonifacio}, {Caffau},
  {Carlberg}, {C{\^o}t{\'e}}, {Fouesneau}, {Fran{\c c}ois}, {Franke},
  {Gonz{\'a}lez Hern{\'a}ndez}, {Gwyn}, {Hill}, {Ibata}, {Jablonka},
  {Longeard}, {McConnachie}, {Navarro}, {S{\'a}nchez-Janssen}, {Tolstoy}, \&
  {Venn}}]{Starkenburg2017}
{Starkenburg}, E., {Martin}, N., {Youakim}, K., {et~al.} 2017, \mnras, 471,
  2587, \dodoi{10.1093/mnras/stx1068}

\bibitem[{{Sukhbold} {et~al.}(2018){Sukhbold}, {Woosley}, \&
  {Heger}}]{Sukhbold2018}
{Sukhbold}, T., {Woosley}, S.~E., \& {Heger}, A. 2018, \apj, 860, 93,
  \dodoi{10.3847/1538-4357/aac2da}

\bibitem[{{Susa} {et~al.}(2014){Susa}, {Hasegawa}, \& {Tominaga}}]{Susa2014}
{Susa}, H., {Hasegawa}, K., \& {Tominaga}, N. 2014, \apj, 792, 32,
  \dodoi{10.1088/0004-637X/792/1/32}

\bibitem[{{Tegmark} {et~al.}(1997){Tegmark}, {Silk}, {Rees}, {Blanchard},
  {Abel}, \& {Palla}}]{Tegmark1997}
{Tegmark}, M., {Silk}, J., {Rees}, M.~J., {et~al.} 1997, \apj, 474, 1,
  \dodoi{10.1086/303434}

\bibitem[{{Trussler} {et~al.}(2022){Trussler}, {Conselice}, {Adams},
  {Maiolino}, {Nakajima}, {Zackrisson}, \& {Ferreira}}]{Trussler2022}
{Trussler}, J. A.~A., {Conselice}, C.~J., {Adams}, N.~J., {et~al.} 2022, arXiv
  e-prints, arXiv:2211.02038.
\newblock \doarXiv{2211.02038}

\bibitem[{{Turk} {et~al.}(2009){Turk}, {Abel}, \& {O'Shea}}]{Turk2009}
{Turk}, M.~J., {Abel}, T., \& {O'Shea}, B. 2009, Science, 325, 601,
  \dodoi{10.1126/science.1173540}

\bibitem[{{van der Walt} {et~al.}(2011){van der Walt}, {Colbert}, \&
  {Varoquaux}}]{NUMPY}
{van der Walt}, S., {Colbert}, S.~C., \& {Varoquaux}, G. 2011, Computing in
  Science and Engineering, 13, 22, \dodoi{10.1109/MCSE.2011.37}

\bibitem[{{Vladilo} {et~al.}(2011){Vladilo}, {Abate}, {Yin}, {Cescutti}, \&
  {Matteucci}}]{Vladilo2011}
{Vladilo}, G., {Abate}, C., {Yin}, J., {Cescutti}, G., \& {Matteucci}, F. 2011,
  \aap, 530, A33, \dodoi{10.1051/0004-6361/201016330}

\bibitem[{{Vogt} {et~al.}(1994){Vogt}, {Allen}, {Bigelow}, {Bresee}, {Brown},
  {Cantrall}, {Conrad}, {Couture}, {Delaney}, {Epps}, {Hilyard}, {Hilyard},
  {Horn}, {Jern}, {Kanto}, {Keane}, {Kibrick}, {Lewis}, {Osborne},
  {Pardeilhan}, {Pfister}, {Ricketts}, {Robinson}, {Stover}, {Tucker}, {Ward},
  \& {Wei}}]{Vogt1994}
{Vogt}, S.~S., {Allen}, S.~L., {Bigelow}, B.~C., {et~al.} 1994, in Society of
  Photo-Optical Instrumentation Engineers (SPIE) Conference Series, Vol. 2198,
  Instrumentation in Astronomy VIII, ed. D.~L. {Crawford} \& E.~R. {Craine},
  362, \dodoi{10.1117/12.176725}

\bibitem[{{Walker} {et~al.}(2016){Walker}, {Mateo}, {Olszewski}, {Koposov},
  {Belokurov}, {Jethwa}, {Nidever}, {Bonnivard}, {Bailey}, {Bell}, \&
  {Loebman}}]{Walker2016}
{Walker}, M.~G., {Mateo}, M., {Olszewski}, E.~W., {et~al.} 2016, \apj, 819, 53,
  \dodoi{10.3847/0004-637X/819/1/53}

\bibitem[{{Welsh} {et~al.}(2019){Welsh}, {Cooke}, \& {Fumagalli}}]{Welsh2019}
{Welsh}, L., {Cooke}, R., \& {Fumagalli}, M. 2019, \mnras, 487, 3363,
  \dodoi{10.1093/mnras/stz1526}

\bibitem[{{Welsh} {et~al.}(2021){Welsh}, {Cooke}, \& {Fumagalli}}]{Welsh2021}
---. 2021, \mnras, 500, 5214, \dodoi{10.1093/mnras/staa3342}

\bibitem[{{Welsh} {et~al.}(2020){Welsh}, {Cooke}, {Fumagalli}, \&
  {Pettini}}]{Welsh2020}
{Welsh}, L., {Cooke}, R., {Fumagalli}, M., \& {Pettini}, M. 2020, \mnras, 494,
  1411, \dodoi{10.1093/mnras/staa807}

\bibitem[{{Welsh} {et~al.}(2022){Welsh}, {Cooke}, {Fumagalli}, \&
  {Pettini}}]{Welsh2022}
---. 2022, \apj, 929, 158, \dodoi{10.3847/1538-4357/ac4503}

\bibitem[{{Wolfe} {et~al.}(2005){Wolfe}, {Gawiser}, \& {Prochaska}}]{Wolfe2005}
{Wolfe}, A.~M., {Gawiser}, E., \& {Prochaska}, J.~X. 2005, \araa, 43, 861,
  \dodoi{10.1146/annurev.astro.42.053102.133950}

\bibitem[{{Wolfe} {et~al.}(1986){Wolfe}, {Turnshek}, {Smith}, \&
  {Cohen}}]{wolfe1986}
{Wolfe}, A.~M., {Turnshek}, D.~A., {Smith}, H.~E., \& {Cohen}, R.~D. 1986,
  ApJs, 61, 249, \dodoi{10.1086/191114}

\bibitem[{{Woosley}(2017)}]{Woosley2017}
{Woosley}, S.~E. 2017, \apj, 836, 244, \dodoi{10.3847/1538-4357/836/2/244}

\bibitem[{{Woosley} \& {Weaver}(1995)}]{WoosleyWeaver1995}
{Woosley}, S.~E., \& {Weaver}, T.~A. 1995, \apjs, 101, 181,
  \dodoi{10.1086/192237}

\bibitem[{{Zou} {et~al.}(2020){Zou}, {Petitjean}, {Noterdaeme}, {Ledoux},
  {Srianand}, {Jiang}, \& {Krogager}}]{zou2020}
{Zou}, S., {Petitjean}, P., {Noterdaeme}, P., {et~al.} 2020, \apj, 901, 105,
  \dodoi{10.3847/1538-4357/abb092}

\end{thebibliography}
\bibliographystyle{aasjournal}



\end{document}